# A practical review on the measurement tools for cellular adhesion force

Rita Ungai-Salánki [1,2,3] *, Beatrix Peter [2], Tamás Gerecsei [1,2], Norbert Orgovan [1,2], Robert Horvath [2], Bálint Szabó [1,2]

[1]*Department of Biological Physics, Eötvös University, Pázmány Péter sétány 1A, Budapest, H-1117 Hungary*

[2]*Nanobiosensorics Group, Research Centre for Natural Sciences, Institute for Technical Physics and Materials Science, Konkoly Thege M. út 29-33. 1121, Budapest, Hungary*

[3] *CellSorter Company for Innovations, Erdőalja út 174, Budapest, H-1037 Hungary*

## Abstract

Cell–cell and cell–matrix adhesions are fundamental in all multicellular organisms. They play a key role in cellular growth, differentiation, pattern formation and migration. Cell-cell adhesion is substantial in the immune response, pathogen-host interactions, and tumor development. The success of tissue engineering and stem cell implantations strongly depends on the fine control of live cell adhesion on the surface of natural or biomimetic scaffolds. Therefore, the quantitative and precise measurement of the adhesion strength of living cells is critical, not only in basic research but in modern technologies, too. Several techniques have been developed or are under development to quantify cell adhesion. All of them have their pros and cons, which has to be carefully considered before the experiments and interpretation of the recorded data. Current review provides a guide to choose the appropriate technique to answer a specific biological question or to complete a biomedical test by measuring cell adhesion.

**Commonly used units:**

Shear stress: 1 dyn/cm$^2$ = 0.1 Pa, 1 pN/µm$^2$ = 1 Pa

Force: 1 dyn = $10^{-5}$ N

**Abbreviations:**

AFM: atomic force microscopy

BFP: biomembrane force probe

CTFM: cell traction force microscopy

DDS: dimethyldichlorosilane

ECM: extracellular matrix

FluidFM: fluidic force microscopy

FRET: fluorescence resonance energy transfer

GETS: genetically encoded molecular tension sensor

ICAM-1: intercellular adhesion molecule-1

MAT: micropipette aspiration technique

NTA: nitrilotriacetic acid

OT: optical tweezers

PA: polyacrilamide

PCR: polymerase chain reaction

PDMS: poly-dymethylsiloxane

PLL-g-PEG: poly (L-lysine)-grafted-poly(ethylene glycol)

PMMA: polymethylmethacrylate

PPFC: parallel plate flow chamber

QCM: quartz crystal microbalance

QPD: quadrant photodiode

RBC: red blood cell

RFC: radial flow chamber

RGD: arginyl-glycyl-aspartic acid

RPM: revolutions per minute

SAM: self-assembled monolayer

$S_c$: critical separation pressure

SCFS: single cell force spectroscopy

SPR: surface plasmon resonance

SPT: step-pressure technique

* Corresponding author: salanki.rita@gmail.com

## Table of contents

# Introduction

Cell adhesion [1] is the ability of a cell to stick to another cell or to the extracellular matrix (ECM). Most mammalian cell types are anchorage-dependent and attach firmly to their environment [2]. Cell adhesion mediated by cell surface receptor molecules, such as integrins [3],[4], cadherins [5],[6], selectins, and members of the immunoglobulin superfamily is a fundamental phenomenon [7] vital for both multi and single cellular organisms (Fig.1). Upon binding their extracellular ligand (such as fibronectin, vitronectin or collagen), integrin molecules cluster to form focal adhesions, complexes containing structural and signalling molecules crucial to the adhesion process [8]. Best-characterized adhesions are the ‘classical’ focal adhesions (also termed focal contacts) and variants including fibrillar adhesions, focal complexes and podosomes [9]. Focal adhesions mediate strong adhesion to the substrate, and they anchor bundles of actin microfilaments through a plaque that consists of many different proteins [9]. Fibrillar adhesions are associated with ECM fibrils. Focal complexes are present mainly at the edges of the lamellipodium. Podosomes contain typical focal contact proteins — such as vinculin and paxillin and they are found in various malignant cells [9].

Communication between cells and the ECM is critically influenced by the mechanical properties of cell surface receptor-ligand interactions. Integrin binding forces were measured in intact cells by atomic force microscopy (AFM) for several RGD containing (Arg-Gly-Asp) ligands and ranged from 32 to 97 pN. The context of the RGD sequence within the ligand protein has considerable influence upon the final binding force [10]. Lee and Marchant reported that the binding force of the single molecular interaction between the RGD-ligand and the integrin was 90 pN [11].

Mechanical forces are central to the functioning of living cells. Cell adhesion is known to be closely related to the actin cytoskeleton, of which organization is crucial in determining the structural and mechanical properties of cells. While significant experimental progress has been made to measure the forces generated by cells, interpretation of these experimental results poses a challenge. Measurements can only be fully interpreted in the light of biophysical models of cell mechanics. However, even a single cell is so complex that theoretical models can usually describe only a few aspects of its behaviour, e.g., contractility considering some cellular components such as stress fibers [12]. In cell cultures on stiff substrates, the actin cytoskeleton tends to be organized into stress fibers, bundles of actin filaments tensed by myosin II molecular motors. Stress fibers usually end in focal adhesions: integrin-based adhesion contacts at the cell membrane. The coupling between mechanics and biochemistry should also be considered in the model [13]. Some models are inspired by architecture like the the so-called tensegrity model [14] of the cytoskeleton [15]. Adhesion-dependent cells sense the mechanical properties of their environment by mechanotransductory processes through focal adhesions. Simplified models applying springs replacing the complex architecture and biochemistry of the cell can be successful to explain certain experimental results [16] A two-spring model [17] showed that the stiffer the environment, the stronger force is built up at focal adhesions by molecular motors interacting with the actin filaments.

Dynamic control of cell adhesion is indispensable to the developing embryo [7], to the immune system [18], and critical in the metastasis of tumors [19],[20],[21] or in the successful implantation of a prosthesis. In case of inflammation, several successive steps of the immune cell adhesion are known [18]. Although the molecular background of this particular process is partially explored there is a lot more to discover about cellular adhesion in general.

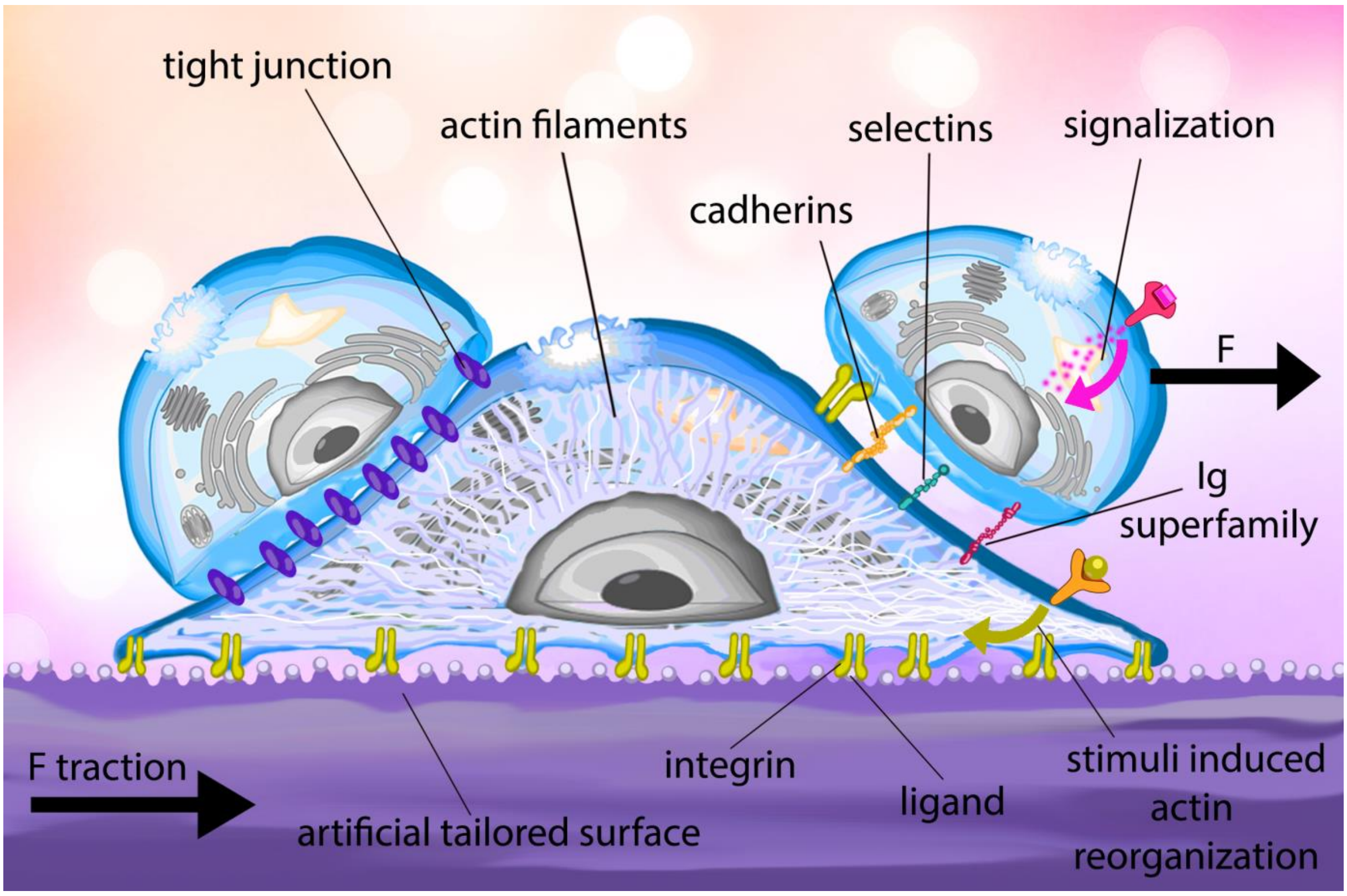

***Fig.1*** *Basic processes of cell adhesion: the artificially tailored surface contains ligands that bind to the integrin receptors found in the membrane of cells. Throughout the process of adhesion, the actin filament structure of the cell is reorganized and a traction force is generated in the substrate. The cytoskeletal reorganization is also regulated by external stimuli. After adhesion to the surface, the cell can interact with other cells through membrane*

*proteins such as cadherins, selectins and members of the Ig superfamily. In tissues a variety of multiprotein complexes called cell junctions (e.g. tight junctions), can form between cells to promote intercellular communication and mechanical stability.*

***In vitro* environments to study cell adhesion.** Both biochemical techniques for tailoring large surfaces on the molecular level and methods capable of the high-throughput high sensitivity measurement of cell adhesion [22],[23],[24],[25] have emerged in the past few years. Appropriate masking (blocking) of the surface area not covered by specific biomolecules has central importance to get rid of non-specific adhesion of cells on, e.g., glass or plastic substrates. For this the protein repellent PLL-g-PEG synthetic polymer [26],[27],[28],[29] proved to be exceptionally useful. Grafting the RGD motif of three amino acid residues into the PEG side chains results in a passivated surface designed for cell adhesion exclusively via RGD-binding cell surface receptors [30],[31]. For investigating cell adhesion mediated by functional proteins (engineered with a His tag) NTA-functionalized PLL-g-PEG [32] offers a versatile possibility. Recently, a 10-fold increase in surface passivation efficiency could be achieved over PEG using the Tween-20 molecule on DDS surfaces [33].

Recent studies demonstrated that a compact and oriented flagellin protein layer - mimicking the surface of bacterial flagellum - has excellent cell repellent capabilities[34],[35],[36],[37]. By genetically engineering the molecule, peptid sequences (for example RGD motif) can be incorporated into this cell repellent layer facilitating receptor specific cellular adhesion. Flagellins can be produced in bacteria and are available in large quantities, opening up an interesting new direction in developing biomimetic engineered surfaces.

**Tools to measure cell adherence**. A number of different techniques can be applied to measure cell adhesion force [38] (Fig.2). Many of them, including the simple washing assay [39], the spinning disk technique [40] and flow chambers [41] are based on the hydrodynamic shear flow removing cells from the surface [42]. However, these techniques do not enable single cell targeting, and cell shape has a strong impact on the shear force making difficult to calculate the exact adhesion force. Furthermore, only weakly adhered cells can be probed due to the technically limited magnitude of shear stress.

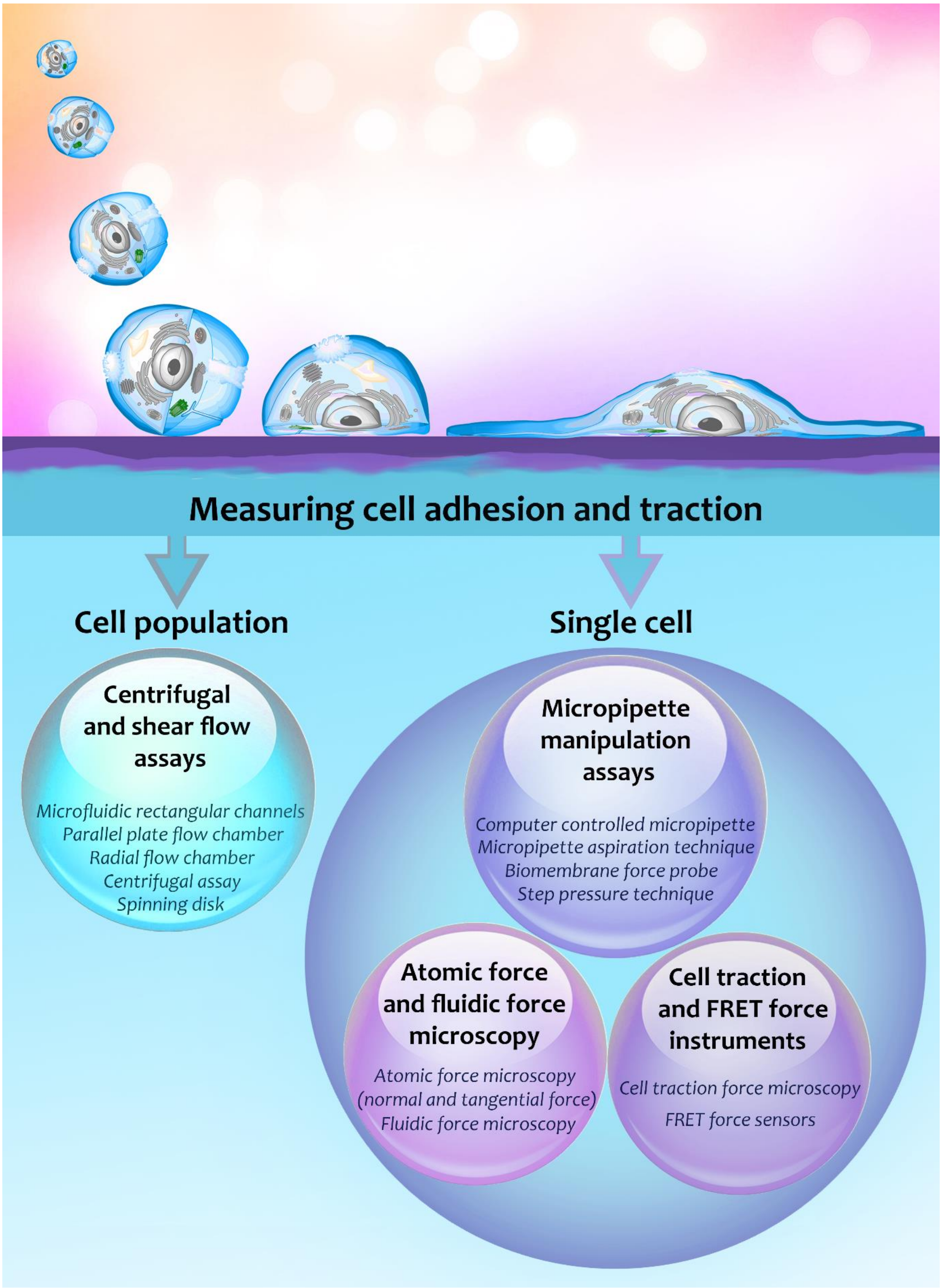

***Fig.2*** *Classification of cell adhesion measurement techniques. Population based assays rely on removing the cells from a surface by hydrodynamic sheer flow. The adhesion force can be calculated from the applied flow rate,*

*however the extracted value depends on other factors as well, such as cell shape. Single cell based techniques can be further divided into categories depending on their principle of operation. Micropipette manipulation assays apply a glass micropipette to aspirate cells from the surface and calculate the adhesion force using hydrodynamic simulations. Atomic force microscopy combined with chemical tip modifications or a nanofluidic channel can also be used to investigate cell adhesion. Other methods include FRET sensors that apply genetically modified cells expressing proteins whose fluorescent signal depends on mechanical tension, as well as force traction measurements capable of detecting forces generated in the substrate by the adhering cell. Data originating from population and single cell based methods differ fundamentally as the population methods are measuring an average over thousands of cells, while single cell measurements can explore differences in heterogeneous populations.*

To directly probe the adhesion force of single cells, cytodetachment with an AFM tip [2],[43],[44] or micropipette aspiration [45],[46],[47] can be chosen as a tool. Both of them are inherently low throughput methods: typically 5-10 cells can be measured in a day. A modified AFM applying vacuum on cells with a fluidic micro-channel (FluidFM) [48] eliminates the painful AFM cantilever chemistry. In a FluidFM, a cantilever can be used for about 10 cells, thus throughput is increased by a factor of 10 compared to conventional AFM. Additionally, the force range is enlarged to reach the μN regime.

Automated glass micropipette equipped onto an inverted microscope can probe the adhesion force of single cells with a relatively high throughput: hundreds of cells can be measured in a ~30 min experiment. Cells can be automatically selected on the basis of computer vision. Single viable cells picked up by the micropipette can be further analyzed by, e.g., DNA/RNA sequencing. Hydrodynamic lifting force acting on cells positioned under the micropipette is calculated from off-line computer simulations.

Cell traction force microscopy (CTFM**)** [49] enables the force (mechanical stress) field generated by single cells on an elastic substrate to be reconstructed with a resolution of ~1-10 μm. Since its introduction in 1996, CTFM has undergone impressive developments. However, the computation of the traction filed is challenging and strongly depends on the mathematical model applied. Thus the derivation of the cellular traction force is not straightforward.

Molecular force sensors [50] inside cells based on Förster resonance energy transfer (FRET) offer a brilliant tool for the high resolution mapping of force distribution inside live cells. The FRET sensor is incorporated into a specific protein of the cytoskeleton in a transgene cell type. The extra gene is constructed to express the two fluorescent proteins separated by an elastic linker region. When the elastic linker region expands as a spring pulled by a mechanic force, the distance of the fluorescent proteins, i.e., the FRET donor and acceptor changes, and can be detected in the fluorescent signal.

An interesting alternative to measure cellular adhesion with high sensitivity and high temporal resolution is the application of novel evanescent field based optical biosensors [22],[51],[52],[53]. Here, the biosensor signal is directly proportional to the cell-substratum contact area and also correlates with the number of biomolecules in the adhesion contact zone, presumably proportional to the strength of adhesion [51]. These label free methods can readily monitor not only the strength but the real-time kinetics of cellular adhesion [22],[54]. But they are indirect, and thus the calculation of the adhesion force is not straightforward; an open problem not yet solved. However, they have clear advantages when the biomolecular adsorption, prior to the adhesion of living cells, has to be also recorded in a single experiment [55].

Surface plasmon resonance (SPR) microscopy can be used as an alternative method for visualizing and quantifying cell/substrate contacts of living cells [56]. SPR, the most well-known evanescent field based optical method is usually used in in biosensor setups to monitor molecular interactions such as receptor/ligand binding in real time. However, SPR has not been demonstrated to measure adhesion forces directly or indirectly.

Quartz crystal microbalance (QCM) with dissipation [57],[58],[59] is a simple, high-resolution mass sensing technique allowing *in situ* real-time measurements of mass and viscoelasticity changes through the resonance frequency (f) and energy dissipation (D) during the cell adhesion process. It can monitor morphological and cytoskeletal changes of surface adherent cells in a non-invasive way [60]. Nevertheless, QCM cannot measure the adhesion force.

Sensitive monitoring of the adhesion process of a large number of single cells on surfaces precisely decorated with biomolecules is expected to produce information on ligand binding, receptor function and signaling pathways with a quality and magnitude used to be unattainable with former cell adhesion assays. Exploring the many faces of molecular scale (nanoscale) direct physical interactions between cells and their environment is a fundamental goal of biophysics, which now seems to be accessible.

Below we divided the available techniques into two categories: population methods and single-cell approaches (Fig.2). Whereas population methods provide cell adhesion data with good statistics, targeting and probing single cells offer a direct and more sensitive measurement of cell adhesion [61]. Inherent heterogeneity of cell populations can be also explored by single cell measurements.

# 1. Centrifugal and shear flow assays

Adhesion force of a cell on a substrate can be probed with simple cell detachment assays like centrifugal or shear flow chambers based on a centrifugal or hydrodynamic shear force to remove weakly attached cells from the substrate [62],[63]. While the applied centrifugal force can act both in a direction normal and parallel to the surface onto cells are attached, shear force is always parallel to the surface. Cells adhered to the surface weaker than the centrifugal or shear force are removed. The exact value of the shear force is difficult to control and therefore ill-defined [2] as it sensitively depends on parameters such as cell size and shape [43]. Although simple washing assays allow the identification of key adhesion components, they provide only qualitative adhesion data. Hydrodynamic shear flow assay applying a well-controlled shear stress to adherent cells is the most common way to quantify cell adhesion. Most widespread methods are the spinning disk, radial flow chambers and parallel plate flow chambers (Table 1).

## 1.1 Centrifugal assay

Cell adhesion strength can be determined using a centrifugal cell detachment assay as described by Chu et. al.(1994) [64] introduced earlier by McClay et al. (1981) [65] and Lotz et al. (1989) [66]. Centrifugal assay employs standard laboratory centrifuges to apply forces on cells adhered to a substrate. Briefly, substrates for cell adhesion are glued to the bottom of a 24-well plate. At the end of the incubation, the wells are filled with medium and covered with sealing tape to avoid medium loss and air bubbles. Then the plates are inverted and spun in a swing-bucket centrifuge for 10 minutes (at 25°C, 800g) [31],[67] (Fig.3). After centrifugation, the number of cells is quantified manually. This assay measures the average response of a cell population and typically examines the fraction of cells that remain adhered to the surface after centrifugation.

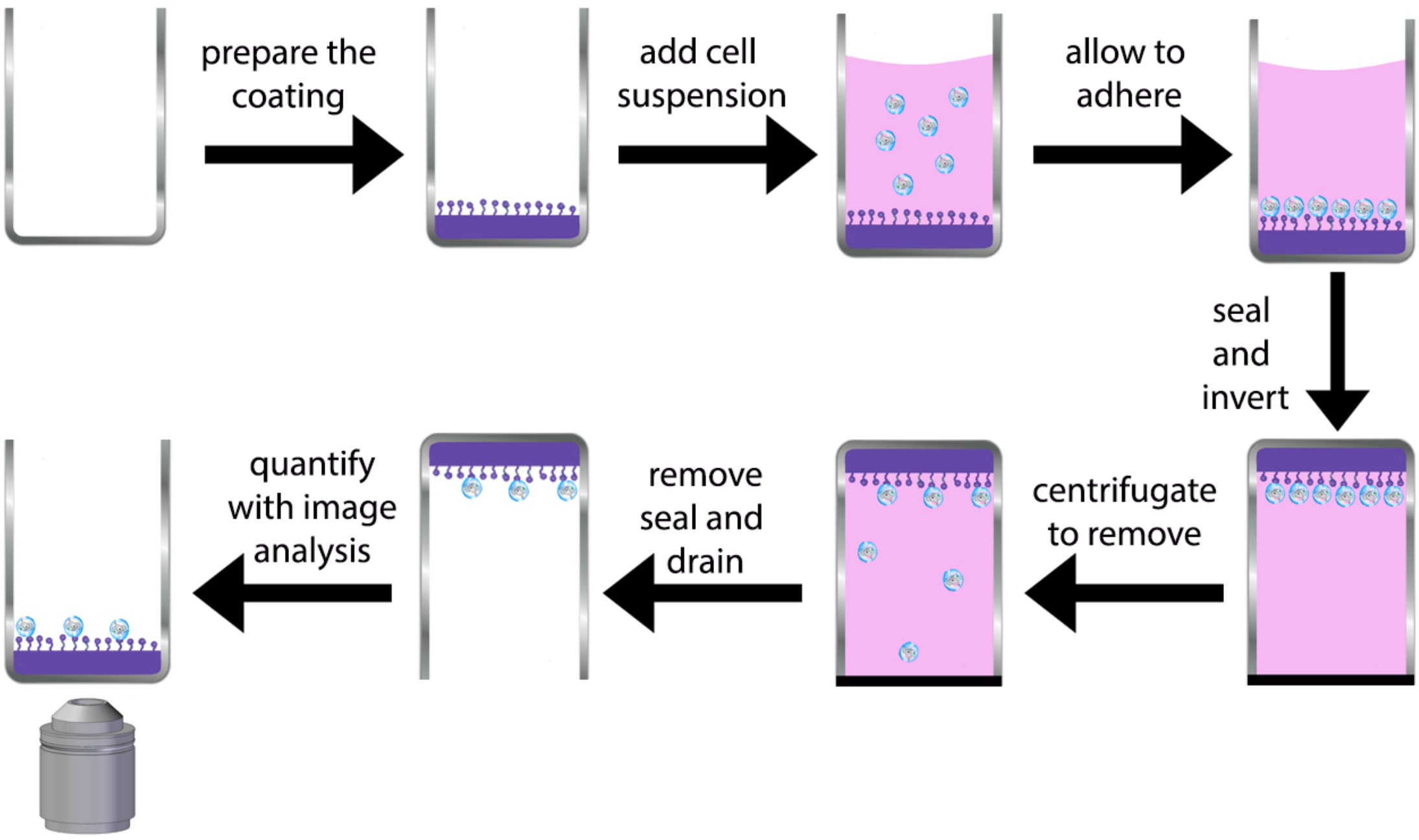

***Fig. 3*** *Steps followed in the centrifugal assay in the study of Chu and co-workers* [64]. *A substrate for cell-adhesion is prepared in a 24-well plate then filled with cell suspension. Cells are let to adhere in an incubator, then the well is sealed airtight and placed in a centrifuge with an opposite orientation. As the centrifugation begins, the cells are affected by a detaching force in the range of 1-100 nN depending on the used acceleration. The fraction of cells detached from the surface can be quantified using optical microscopy. The presented method is simple and cost-efficient, however the force range is limited by the centrifuge's properties. Even with ultracentrifuges reaching 100,000 g in acceleration. Only weakly adhering cells can be studied. Furthermore, the cell shape is altered significantly by the high acceleration endured by the cells during centrifugation.*

*a. Number of cells in an experiment*

Typical cell number is $10^5$ cell [68].

*b. Typical range of centrifugal force*

This assay applies controlled detachment forces to a large population of adherent cells [69].To determine the adhesion strength, the number of cells is quantified before and after applying the centrifugal force [42]. The force in the centrifuge is calculated as follows:

$$\mathbf{F = G \times V_{cell} \times (\rho_{cell} - \rho_{medium}),} \qquad (1)$$

where $F$ is the force exerted on the cell, $G$ is the centrifugal acceleration, $V_{cell}$ is the cell volume, $\rho_{cell}$ is the density of the cell (typically 1.07 g/cm$^3$), $\rho_{medium}$ is the density of the medium (typically 1.00 g/cm$^3$) [31],[42]. Typical G values fall in the range of 20-1,000g and the forces on an individual cell is 1-2,000 pN  When using an ultracentrifuge, adhesion force can be measured up to 100 nN [70],[71] with a maximum acceleration of 110,000 g. However, before cells detach due to such a high acceleration, cell shape changes dramatically similarly to shear force induced shape alteration.

*c. Experiment duration*

The duration of load application in centrifuge tests typically range from 5 to 10 min [31],[42],[69],[72].

*d. Advantages*

Simplicity. It does not require specialized equipment: the method is widely accessible [42].

*e. Disadvantages*

Normally restricted to the investigation of weakly-adhered cells [42].

*f. Main applications*

Adhesion of MC3T3-E1 cells on self-assembled monolayers (SAMs) as a function of fibronectin density [42],[69] . Cell-cell adhesion mediated by E-cadherin [68]. Cell adhesion to specific RGD-modified substrates over a wide range of forces [31]. Strength of bacterium-receptor interactions [73].

### 1.2 Spinning disk

The apparatus applies a disk spinning in a large volume of fluid. The cell adhesive surface is mounted onto a rotating circular stage, which produces a fluid flow in the chamber [74] . Fluid flow over the cells on the disk creates a detachment force [75],  which is calculated from the properties of the buffer and the rotational speed (Fig.4). At the axis of rotation the detachment force is zero and it increases linearly with the distance from the axis. This fewer cells will remain near the edge of the disk than closer to the axis. After spinning, the remaining adherent cells are fixed, stained, and counted.

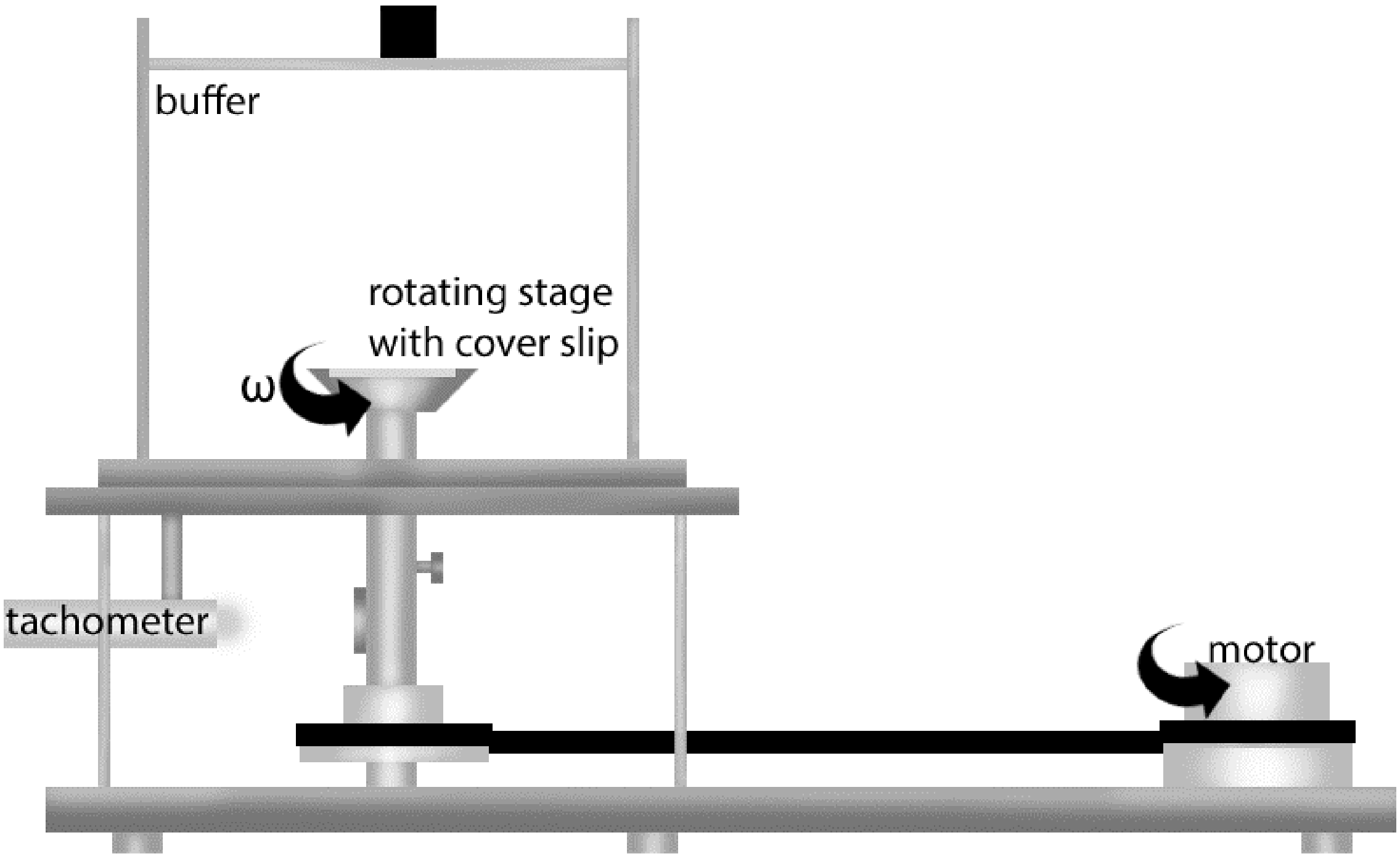

***Fig.4*** *Spinning disk arrangement in the experiment of Wertheim et al* [74]. *Cells are allowed to settle onto a fibronectin-coated matrix atop the stage. As the stage is rotated, the spinning stage inside the fluid is generating a sheer flow on the surface. The stress caused by this flow is linearly increasing from the center of the disk to the edge, causing a number of cells to detach in the function of distance from the center. The distribution of cells that remain on the stage after spinning can be quantified by optical microscopy. Because of the linearly changing shear stress on the surface, a range of forces can be studied at the same time. The disadvantage of this method is common in all shear flow based techniques namely the dependence of the measured detachment force on cell shape. This particular construction also makes it impossible to optically observe the cells during measurement.*

*a. Number of cells in an experiment*

400 cells/mm$^2$ [76] are uniformly seeded onto coverslips.

*b. Typical range of shear stress*

Detachment force is proportional to the hydrodynamic wall shear stress, $\tau$ (shear force/area)

$$\tau = 0.800\, r \sqrt{\rho \mu \omega^3} \qquad (2)$$

where $r$ is the radial position from the disk center, $\rho$ and $\mu$ are fluid density and viscosity, respectively and $\omega$ is the angular speed (rad/s). Typical frequency range: 500-3000 RPM [42].The shear force increases linearly with the radius from the center of the disk, thus it can generate a wide range of forces at the same time.

A value of 200 Pa shear stress was reported [77]. The range of detachment forces is (0-10 $Pa^2$) [74].

*c. Experiment duration*

Disks were usually spun for 5- 10 min at constant speed [76],[77].

*d. Advantages*

The key advantage of the method is the shear stress increases linearly with the radial position on the disk allowing a wide range of forces to be applied at the same time [42],[74]. Spinning disk method can generate high shear stresses and detach relatively strongly-adhered cells.

*e. Disadvantages*

Hydrodynamic shear force depends on cell shape [75]. It is not compatible with simultaneous microscopic imaging [61].

*f. Main applications*

Binding strength between cells and Fn-coated micropattern islands [40]. Modell system: erythroleukemia cell line, expressing a single fibronectin receptor, integrin α5β1. Cell detachment profile and mean detachment force measurements [75],[77]. Bacterial spore adhesion [78].

### 1.3 Radial flow chamber

In the axisymmetric flow geometry of radial flow chambers (RFC), a wide range of radially dependent shear stress is applied to the adherent cells in a single experiment [79]. A design of RFC by Stone (1993) [80] employs a chamber on an inverted microscope, allowing *in situ* observation of adherent cells exposed to shear stress. The surface is pre-incubated, then the cells are introduced into the RFC by a syringe and allowed to adhere to the surfaces for a given time (~30-60 min), prior to applying the shear.  Fluid [79],[81] flows radially between the surface with the cells and a glass slide at a constant volumetric flow rate for ~5 min (Fig.5). The chamber has a 200 µm gap between the surfaces. Cells are removed, where the shear force exceeds the adhesive force of cells [80] .

The mean velocity of the fluid and hence the shear stress, decreases with increasing radial position, *r*. For the radial flow, the Reynolds number is inversely proportional to radial position:

$$\mathbf{Re}=\frac{2\rho Q}{\pi\mu r} \tag{3}$$

where $Q$ is the flow rate, $\rho$ and $\mu$ are the density and viscosity of the fluid, respectively. To ensure laminar flow, the critical upper limit to $Re$ is *2000*.

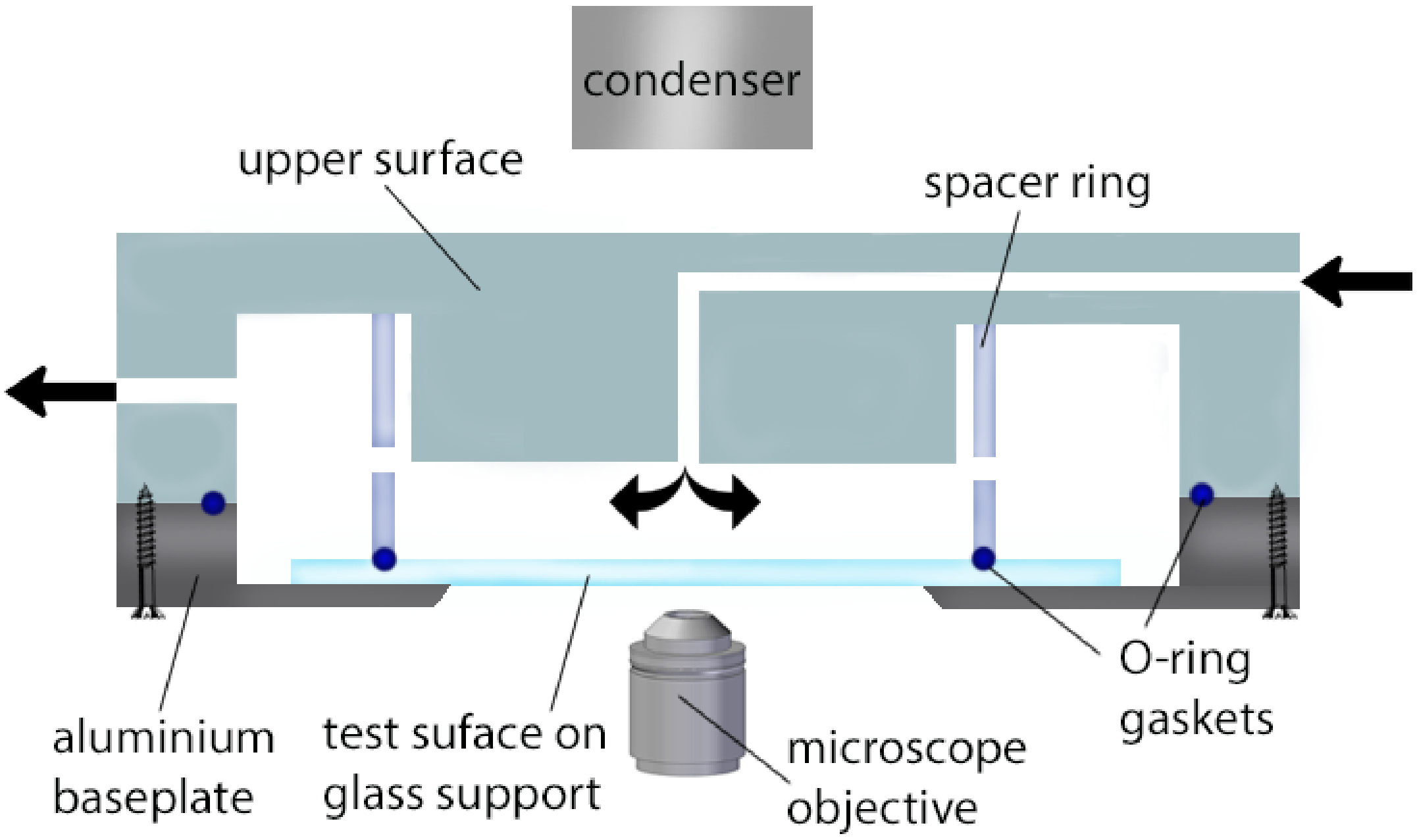

***Fig.5*** *Schematic illustration of the radial flow chamber based on the publication of Goldstein et al. (1998)* [81]. *The cells are seeded on the glass support, exposed to a laminar flow coming from above. The flow rate determines the shear stress on the cells up to the point where the flow would become turbulent. Much like in the spinning disk case, the stress is linearly changing with the position relative to the center of the dish but in this case the maximum*

*falls to the center and decreases as we approach to the edge. This construction can be directly mounted on an inverted microscope thus enabling the optical monitoring of the detachment process.*

*a. Number of cells in an experiment*

$10^5$ cells / ml added to a 60-mm-diameter dish. Seeding density: ~40 cells/ $mm^2$ [79],[80].
Cells are plated at an approximate density of $10.25 \times 10^4$ cells /mL (~$8.34 \times 10^4$ cells/$cm^2$, ~$2.64 \times 10^6$ cells/surface) [82].

*b. Typical range of shear stress*

Cells were allowed to attach to fibronectin-coated substrates for 30 min and then subjected to a spatially dependent range of shear stress for 5 min (28–220 dyn/$cm^2$) [83].
After 75 to 90 min, at a shear stress of 350 dynes/$cm^2$, more than 50% of the spread cells are detached from the surface. Cells with higher spreading areas stay longer at the glass surface [84].

*c. Experiment duration*

Cells are incubated for 30-60 min. Shear stress is applied for ~ 5 min [79],[81],[83],[84].

*d. Advantages*

Broad range of the shear stress simultaneously [81]. If the RFC is mounted on an automated microscope, *in situ* observation of cell detachment can be monitored under the hydrodynamic shear stress [79],[83]. It is allowed the application of uniform stress fields to cells while enabling visualization of the cells via microscopy [61].

*e. Disadvantages*

These chambers are unable to generate stresses high enough to detach well-spread cells under laminar flow conditions [61]. Shear stress limitations due to the upper limit of Reynolds number to avoid turbulence [81]. Complex fluidics needed [85].

*f. Main applications*

*In situ* detachment of 3T3 murine fibroblasts from substrata of SAMs of dodecanethiolate [79],[81] and fibronectin. Cell deformations (elongation) after applying the shear stress [79],[86]. Adhesive nature of the modified substrates [82]. Bacterial adhesion experiments [87].

### 1.4 Parallel plate flow chamber

In the parallel plate flow chamber (PPFC) (Fig.6) the flow is precisely defined to produce a uniform shear stress [84],[88],[89],[90]. Shear stress τ (dyne/$cm^2$) is calculated from the following equation:

$$\tau = 6Q\,\mu / w\,h^2 \qquad (4)$$

where $Q$ is the volumetric flow rate ($cm^3$/s), $\mu$ is the viscosity of the medium, $w$ is the width of the chamber and h is the height of the chamber. Usually, a peristaltic pump is used to adjust the required flow rate [90].

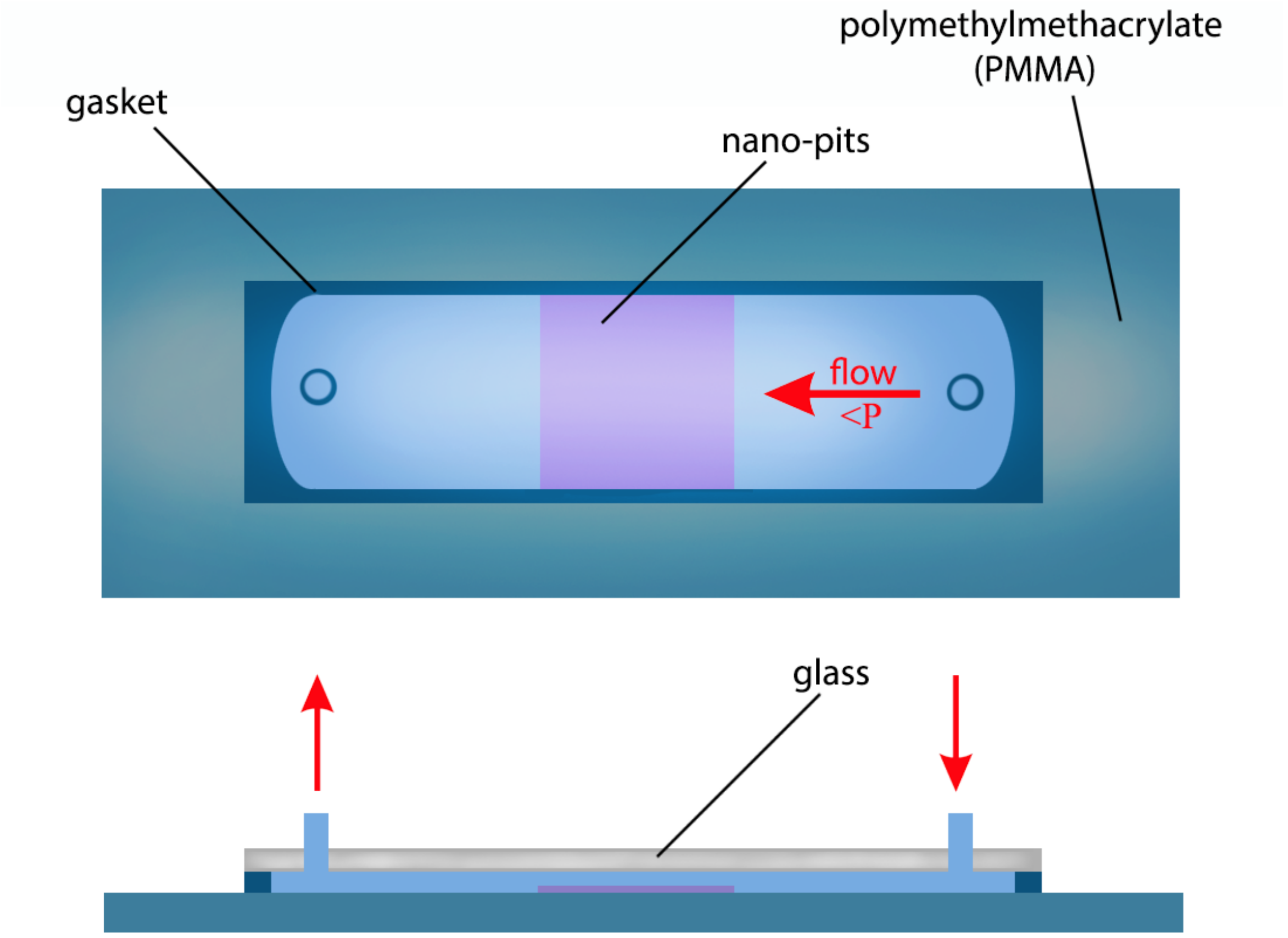

***Fig.6*** *Schematic illustrations of a parallel plate flow chamber used to characterize the effect of nanotopology on cell adhesion based on the publication of Martines et al.(2004)* [91]. *A polymethylmethacrylate (PMMA) sample (turquoise) was patterned with 100 nm nanopits, then a glass slide (light blue) was attached to it with a thermoplasitc gasket (dark blue) in between. The channels in the PMMA were designed to create a laminar flow within the chamber. The inlet and outlet for the solution is provided by circular holes in the glass slide on the top of the device. The adhesion strength can be measured by monitoring the attached cell's density as a function of the flow in the channels. The authors have successfully shown the adhesion reducing property of the nanopatterned surface topology.*

*a. Number of cells in an experiment*

This technique can be used to monitor the removal of a population of cells from a surface. Seeding concentration: $10^4$ -$10^6$ cells /ml resulting in 100-1,000 cells/mm$^2$ [88],[91],[92]. However, normally only few hundred cells are analyzed.

*b. Typical range of shear stress*

1-25 Pa [89],[90],[92],[93]. Cells can be exposed to either a single shear stress or incremental shear stress levels [89]. Also the effect of increasing pulsatile flow on cell adhesion can be monitored [93].

*c. Experiment duration*

Time elapsing after seeding cells on the surface, and before applying the shear stress: 1-24 hours [88]. Duration of shear stress can be several hours [90]. In the incremental shear studies shear stress can be increased in the time scale of minutes (4-15 min) [89],[93]. In ref [92] cells were exposed to a range of shear stress levels each applied for only 2-10 sec.

*d. Advantages*

Parallel plate flow chamber can be utilized widely due to its simple well-characterized flow regime [89]. The chamber can be mounted onto a microscope, thus the behavior of the cells as a response to the shear stress can be observed *in situ*. Parallel flow chambers can provide a precisely controlled, uniform shear stress with dynamic, well-defined shear regulation [94]. In this chamber the motion of a sphere can be easily calculated. Cell trajectories, speed and adhesion process can be monitored [91]. It is a frequently applied design because of its simplicity. It can be employed on a broad range of surfaces, such as silicone rubber, dental enamel and metals [95].

*e. Disadvantages*

The most parallel plate flow chamber have a height >200 μm [61]. This sets a practical limit on the shear stress value considering the normally available range of flow rate or overpressure.

*f. Main applications*

Channels can be used to determine the strength of adhesion of fibroblast cells to various substrates, with different wettabilities [89]. Cell retention, morphology and migration as a function of flow rate and the influence of adhesion time were also studied [88]. Endothelial cell adherence onto polymer surfaces with different hydrophilicity [90]. Investigation of nano-patterned surfaces [91]. Development of flow circuits, in which the behavior of cells can be continuously monitored on a microscope [88]. The flow circuit usually consists of three parts: a flow loop, a heating system, and an image analysis component [88],[89]. Adhesion of bacterial and yeast strains to a PEO (Poly(ethylene oxide))-brush covalently attached to glass [96]. Measurement of the attachment and detachment rates of Escherichia coli to and from a glass surface [97]. Nonspecific surface attachment of hydrophobic yeast cells [98].

### 1.5 Microfluidic rectangular channels

This technique is similar to the parallel plate flow chamber, but the key advantage is the small characteristic dimension of the flow channel (h < 100 μm) that allows substantially higher shear stresses to be generated under laminar flow conditions [61] (Fig.7). The channels can be readily mounted onto a microscope, and thus cell detachment can be monitored *in situ*. The device is usually constructed from PDMS, because it is quick and simple to use, cheap, transparent, and biocompatible [99],[100].

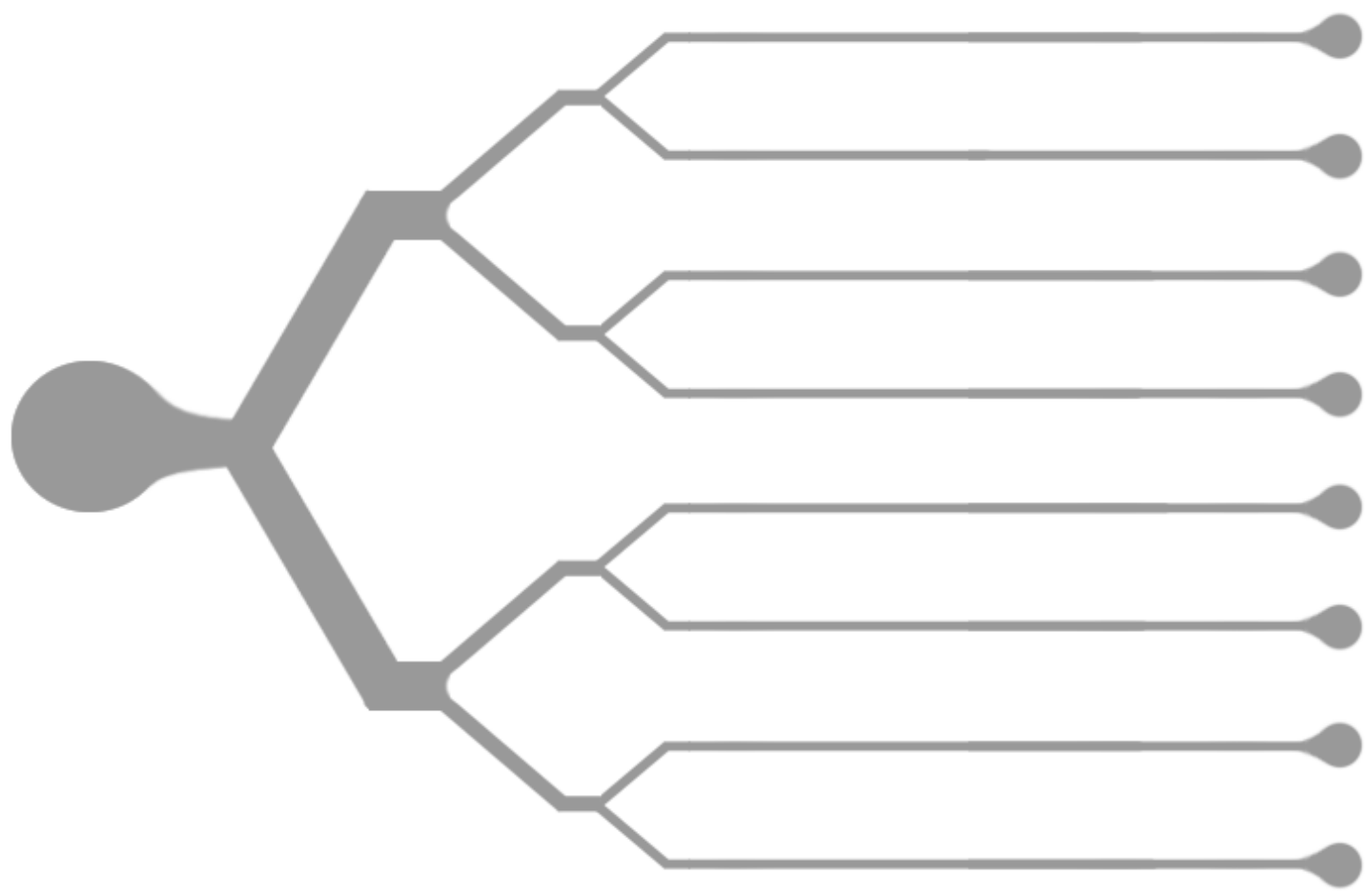

***Fig.7*** *Parallel microfluidic shear device pattern, consisting of eight microchannels used in the study of Young and co-workers* [101]. *The flow is directed into the device through the common inlet on the left. The 516 µm wide and 59 µm high channels are designed to ensure identical flow in all branches. The main advantage of this arrangement is the parallelization of measurements: a fraction of the channels were coated with proteins, while others were used as control during the same experiment. Cell spreading and attachment can be monitored by optical observation of the channels by an inverted microscope.*

*a. Number of cells in an experiment*

Cell suspension density: $10^6$ -$10^7$ cells/ ml Number of analyzed cells: 100 - 1,000. Usually 1-10 microscopic fields of view with 10-100 cells/ field of view [61],[100],[101],[102].

*b. Typical range of shear stress*

50-200 Pa [61]. When the shear stress is time dependent, its rate is usually in the 1-10 Pa/sec range [61]. 1 - 20 Pa [101].

*c. Experiment duration*

Typical experiment duration: 10-15 min [61]. Adhesion strength of cells was investigated in 12 min, while shear stress level was increased step by step in 4 min intervals in ref [101].

*d. Advantages*

This method offers a wide range of forces and relatively high-throughput. The experimental setup is simple, and it can be integrated with other microanalytic modules. Using time-lapse microscopy, cell detachment can be monitored *in situ.* The microfluidic device offers advantages with its small dimensions and ensures laminar flow even at high fluid velocities. The small channel height (h<100 μm) enables higher shear stress than the similar parallel plate flow channel. Thus also the strongly adherent cells (cannot be studied in most conventional assays) can be investigated in microfluidic rectangular channels [61]. The device is usually fabricated from poly-dymethylsiloxane (PDMS) bonded to glass using the rapid technique of soft photolithography [100]. The technique has small sample consumption and it provides a stable and well characterized flow [101],[102]. A parallel microfluidic network of the channels permits the parallel analysis of multiple samples or conditions [101].

*e. Disadvantages*

At the micron scale, channel deformation effect becomes important and must be quantified for predictable assay performance [99]. Cell morphology is affected by the flow: cells become elongated due to the applied shear stress. It shows that the flow can significantly alter the cytoskeleton several minutes before cells detach from the surface [101]. This "side-effect" needs to be thoroughly considered when evaluating experimental results.

*f. Main applications*

Measurement of the adhesion strength of well-spread cells on different surfaces [61]. Investigating the adhesion difference between normal and cancerous epithelial cells on nanostructured polymer surfaces. Adherence of vascular and valvular endothelial cells on different extracellular matrix proteins (fibronectin, collagen I) [101]. Observation of bacterial adhesion [103],[104].

## 2. Micropipette manipulations

The use of micropipette manipulation studies can be traced back to Mitchison and Swann (1954). They first developed and applied a micropipette-based elastimeter to determine the membrane elastic modulus and internal pressure of the unfertilized sea urchin eggs. The method can measure the deformation of a single cell attached to the tip of a micropipette by a precisely controlled vacuum in the micropipette [104]. Ten years later, Rand and Burton (1964) refined this method. They measured the stiffness of the erythrocyte (RBC) membrane and the pressure inside the cell [105],[106]. In the 1970-80s, the micropipette manipulation method was further improved and a number of studies applied it to explore the mechanics of RBCs, white blood cells, endothelial cells, lipid vesicles, and liposomes. These studies imposed either a local force or a suction underpressure on individual cells or liposome surface. In the 1990s, the technique was extended to studies including tumor cell metastasis and the kinetics of cell-cell contact interactions [46]. The common approach is to quantify the adhesion strength by imposing tension to rupture the adhesive contact between two opposite faces. An increasing force is employed to measure the magnitude of the rupture force. Alternatively, a constant force is applied to determine the adhesion lifetime [46].
Micropipette manipulation techniques are summarized in Table 2.

### 2.1 Step-pressure technique (SPT)

Sung et al. (1986) developed the first micropipette based technique, which was used to quantify cell adhesion strength [47]. The system consists of a cylindrical glass micropipette with an internal diameter of a few μm-s and a manometer to control the pressure inside the micropipette with a resolution of 0,01pN/ $\mu m^2$ [43],[107],[108]. One cell is tightly held by micropipette 1 with high suction vacuum, while the second is held by micropipette 2 with lower suction force. First, the two cells are brought into contact, so they can adhere to each other. Then micropipette 2 is pulled away (Fig.8). The suction vacuum in micropipette 2 is increased until the cells are separated from each other. The minimum suction vacuum which leads to the separation of the cells is the critical separation pressure ($S_c$). One of the cells can be replaced with a protein coated substrate to determine the cell- substrate adhesion strength [46].

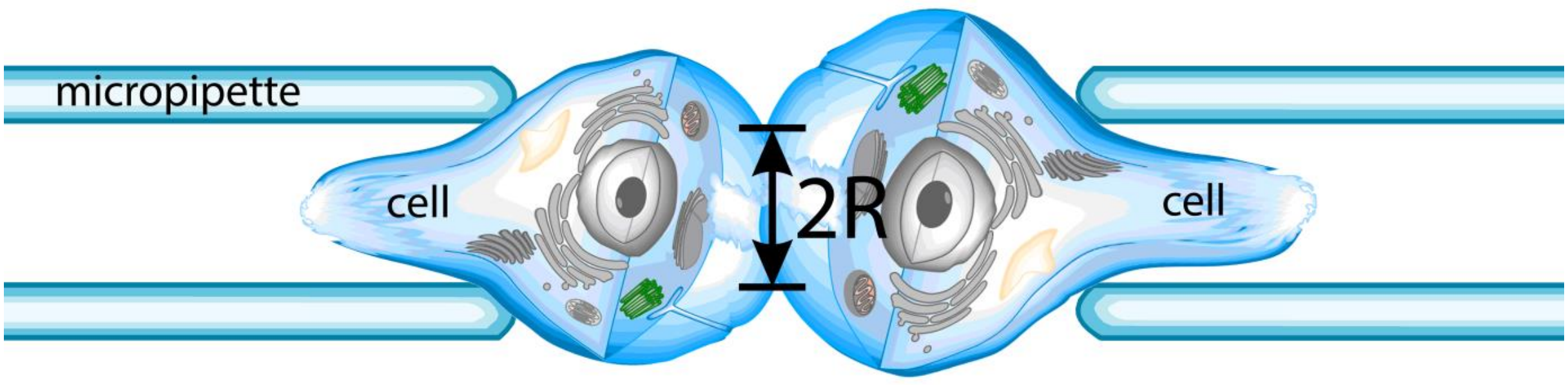

***Fig.8*** *Schematic illustration of step-pressure technique (based on the publication of Shao et al.)* [46]. *Two micropipettes each with a cell sucked into their tips are facing each other as the cell's surfaces are pushed together. After a certain adhesion time, the cell on the left is pulled away. If the moving cell slips out of the micropipette, the suction force was smaller than the adhesion force between the cells. By increasing the suction force, an upper estimate can be determined for the inter-cell adhesion force and energy. This method was the first to be used to measure single cell adhesion by glass micropipettes. This technique allows a precise control of cell contact area and time, however the nonspecific adhesion between the glass tips and the cell membrane must be eliminated. Another disadvantage of any similar method is the dependence of the measured adhesion properties on the speed of the separation of the micropipette tips.*

*a. Number of cells in an experiment*

~10 pairs of cells [109].

*b. Typical force range*

The applied vacuum is in the 10-100 Pa range [109],[110]. Sc =1,5 nN/$\mu m^2$, However, it is difficult to measure the exact contact area between the cells. Separation force is in the 1-10 mN range [46],[47].

*c. Experiment duration*

Cells are brought into contact for ~10 sec [110]. In ref [47] Sung et al. measured the adhesion force between two cells for 120 min. The first measurement started at 12 min and subsequent measurements were taken in every 10 min.

*d. Advantages*

Enables the measurement of the adhesion force between two similar or different single cells *in vitro* [47]. Cell-cell contact area and contact time can be controlled [46],[110].

*e. Disadvantages*

Nonspecific adhesion between the cell and pipette wall has to be much weaker than the overall adhesion to the other cell or to the substrate [46]. Low throughput, manual measurement.

*f. Main applications*

Enables the real time observation of two cells assembling and then losing the contact zone in between them. Cell adhesion force can be measured directly [46].

### 2.2 Biomembrane force probe (BFP)

Evans et al. (1995) described a simple apparatus and procedure for probing weak forces at biological surfaces [111]. The technique applies a force transducer (these are cell-sized membrane capsules such as a vesicle, liposome or an erythrocyte) which is held by a micropipette with a suction force while a micro-bead is attached to the force transducer [46],[47]. The microscopic bead is glued biochemically to the transducer. The bead is coated with the protein of interest in order to interact with the cell (Fig.9). Pulling or pushing force applied to the bead result in membrane deformation of the force transducer [46].

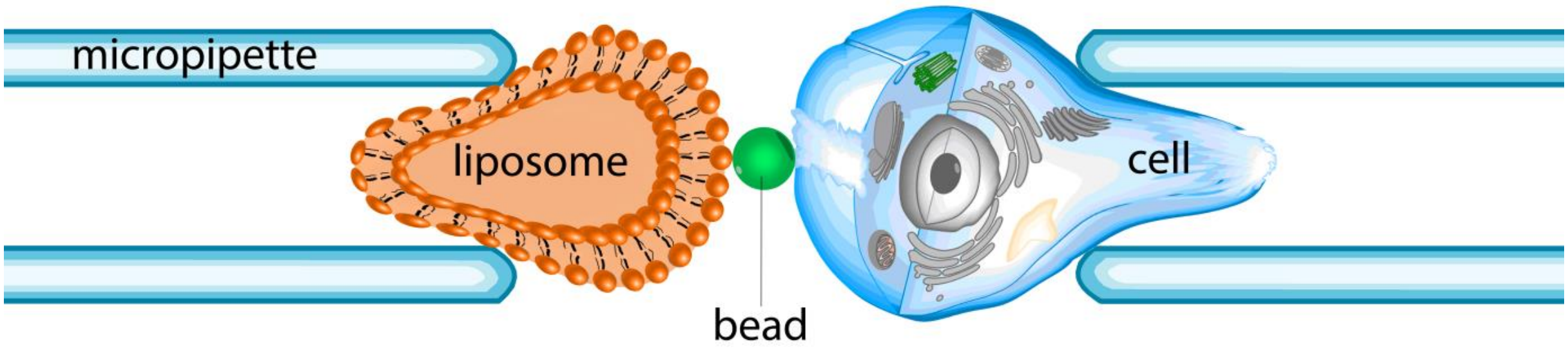

***Fig.9*** *Schematic illustration of biomembrane force probe (Based on the publication of Shao et al. [46]). A liposome is sucked into a micropipette that is biochemically glued to a latex bead. By pushing or pulling the liposome, a force can be exerted on the bond between the cell and the bead. As the latter can be coated with a protein of interest, receptor-ligand bonds can be studied under a controllable loading rate and force. Using this arrangement Evans et al. showed that the rupture force of a ligand-receptor bond depends on the speed at which the force is applied.*

*a. Number of cells in an experiment*

1 cell/experiment.

*b. Typical force range*

Evans et al. (1998) employed two modes of BFP; vertical and horizontal modes. Vertical mode was used to test weak bonds under slow loading, when the force was 0.2-0.5 pN; while the horizontal mode was utilized to study stronger bonds under fast loading (1-10 pN) [112]. The transducer sensitivity is tuned to measure forces from 0.01 to 1,000 pN with a range of loading rates from 0.1pN/s to 100,000 pN/s. Evans et al. have shown that, rupture force of a receptor-ligand bond depends on the loading rate of the force, i.e., how fast the bond is pulled [46],[113],[114],[115],[116],[117],[118].

*c. Experiment duration*

No information.

*d. Advantages*

The loading rate of the force is adjustable [46]. The interference to the cell can be minimized. Gentle measurement minimally altering the cytoskeleton. Weak adhesive bonds can be readily detected. Sensitivity of the technique enables the detection of local activation of cytoskeletal structures [111]. The method has sub-pN force and nanometer scale displacement resolution [46].

*e. Disadvantages*

The resolution of probe movement is not as good as in AFM [111].

*f. Main applications*

It allows the quantification of single molecular bonds [38]. Receptor-ligand binding [46].

## 2.3 Micropipette aspiration technique (MAT)

The technique was developed by Shao and Hochmuth (1996). They applied the method to measure the magnitude of pN forces [46],[107]. Their aim was to create a simple method based on micropipette suction, allowing the measurement of detachment force of one cell from another cell or a solid surface [107] (Fig.10).

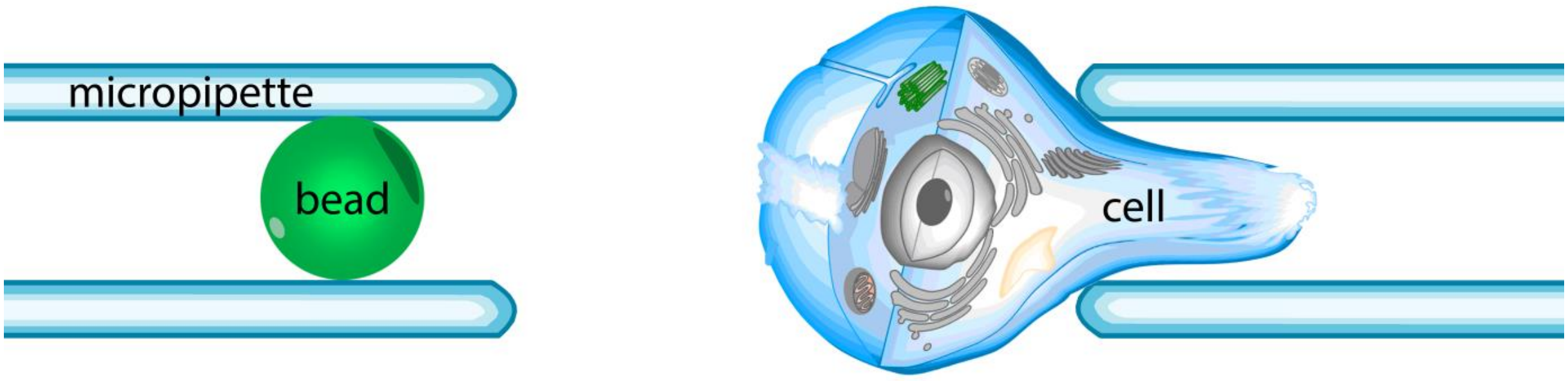

***Fig.10*** *Schematic illustration of micropipette aspiration technique (Based on the publication of Shao et al.)* [46]. *In the micropipette on the left a bead is aspired that can move freely within the glass walls. As the micropipette is pushed to the cell, the functionalized bead surface is brought into contact with the cell membrane using a gentle pressure. After the ligand-receptor bonds are formed, the bead is sucked into the micropipette with a controllable force. The point at which the bead detaches is the adhesion force as measured by this technique. Alternatively, the bead can be replaced with a spherical cell (Spillmann et al.)* [119] *to directly study cell-cell interaction.*

*a. Number of cells in an experiment*

One cell/experiment.

*b. Typical force range*

A spherical object either a cell (e.g. human neutrophil) or a bead is used as a MAT transducer and it can exert forces lower than 10-20 pN [46]. Between the transducer and the pipette wall, a small clearance is necessary to allow free movement of the transducer inside the pipette. A freely sliding cell or bead is used as a force transducer. A positive pressure allows the bead (which is coated with the protein of interest) to contact and adhere to the cell [107]. Then, a constant suction underpressure provides the tensile force to break the adhesive bond. According to the diameter of the pipettes (1-10 μm), the force range is 10 pN-1 nN [45]. The force exerted by the static cell (or bead) can be determined as follows:

$$F=\pi R^2_p \Delta p \qquad (5)$$

where $R_p$ is the radius of the pipette, $\Delta p$ is the suction underpressure [80]. A bead was partially aspirated by a micropipette as described by Shao and Hochmut (1996) [107]. The diameter of the pipette was smaller than the bead, thus the bead was attached to the opening of the pipette. Another micropipette was used to aspirate a neutrophil cell. Diameter of this pipette was slightly smaller or equal to the diameter of the neutrophil. Adhesion strength is defined as the minimum force needed to detach a single cell from its substrate [38]. The technique was used to determine the minimum force (45 pN) necessary to form a membrane tether from neutrophils [107]. The smallest bead (3.2 μm) applied, resulted in a force of a few pN. As a high force limit, hundreds of nN could be exerted [46].

*c. Experiment duration*

Lomakina (2004) hold a bead and a neutrophil in contact for a user-specified length of time (2s and 1 min ) and then separated them [120].

*d. Advantages*

The strength of the MAT is in its simplicity. It can measure the force between cells without attaching cells onto a solid surface. Cell-cell interactions can be studied directly. Interestingly, a spherical cell can be used as a force transducer [46]. A constant localized force can be imposed to monitor the effect of the force on single-bond kinetics [46]. While the low force sensitivity of the MAT is similar to that of optical tweezers, it can exert much higher forces as well [46].

*e. Disadvantages*

One of the cells (or probe) has to fit snugly inside the pipette. Forces smaller than 10-20 pN cannot be measured precisely when the diameter of the pipette is ~10 μm [107]. As the adhesion between the bead-transducer and the pipette wall would affect the measurement, it is important to insure that the bead does not adhere to the pipette wall [46]. Evaporation in the chamber can be a significant technical issue [107].

*f. Main applications*

Viscoelastic properties of soft cells e.g. red blood cells, white blood cells; and more rigid cells, such as endothelial cells can be measured with it [45]. Micropipette suction is a versatile method to study mechanical properties of living cells and to examine the viscous response of solid cells e.g. endothelial cells, chondrocytes [45]. It can measure cell-cell interactions, directly Lomakina et al. [46],[119],[120] investigated interactions between neutrophils and ICAM-1 coated substrates using 2 micropipettes. One of them (stationary pipette) hold the bead and the other hold the neutrophil to manipulate the cell [120].

### 2.4 Computer-controlled micropipette

Computer-controlled micropipettes can manipulate and sort cells in a Petri dish, individually [121],[122] (Fig.11). Cells are selected on the basis of their phase contrast and/or fluorescent images). Sorting is performed by a micropipette with an aperture of 10-70 μm with a sorting speed of 3-4 cell/min. After sorting, single cells are deposited into another Petri dish/multiwall plate/PCR tube or glass cover slip in a minute volume of liquid in the nanoliter-microliter range. Individual cells inside the drops on the glass cover slip can be studied with high resolution, immediately after sorting [123]. The technique is suitable for high-throughput single cell adhesion force measurements by repeating the pick-up process with an increasing vacuum. Adhesion force between individual human white blood cells and specific macromolecules were studied with the technique [124],[125],[126].

Hundreds of cells adhered to specific macromolecules can be measured one by one in a relatively short period of time (~30 min).

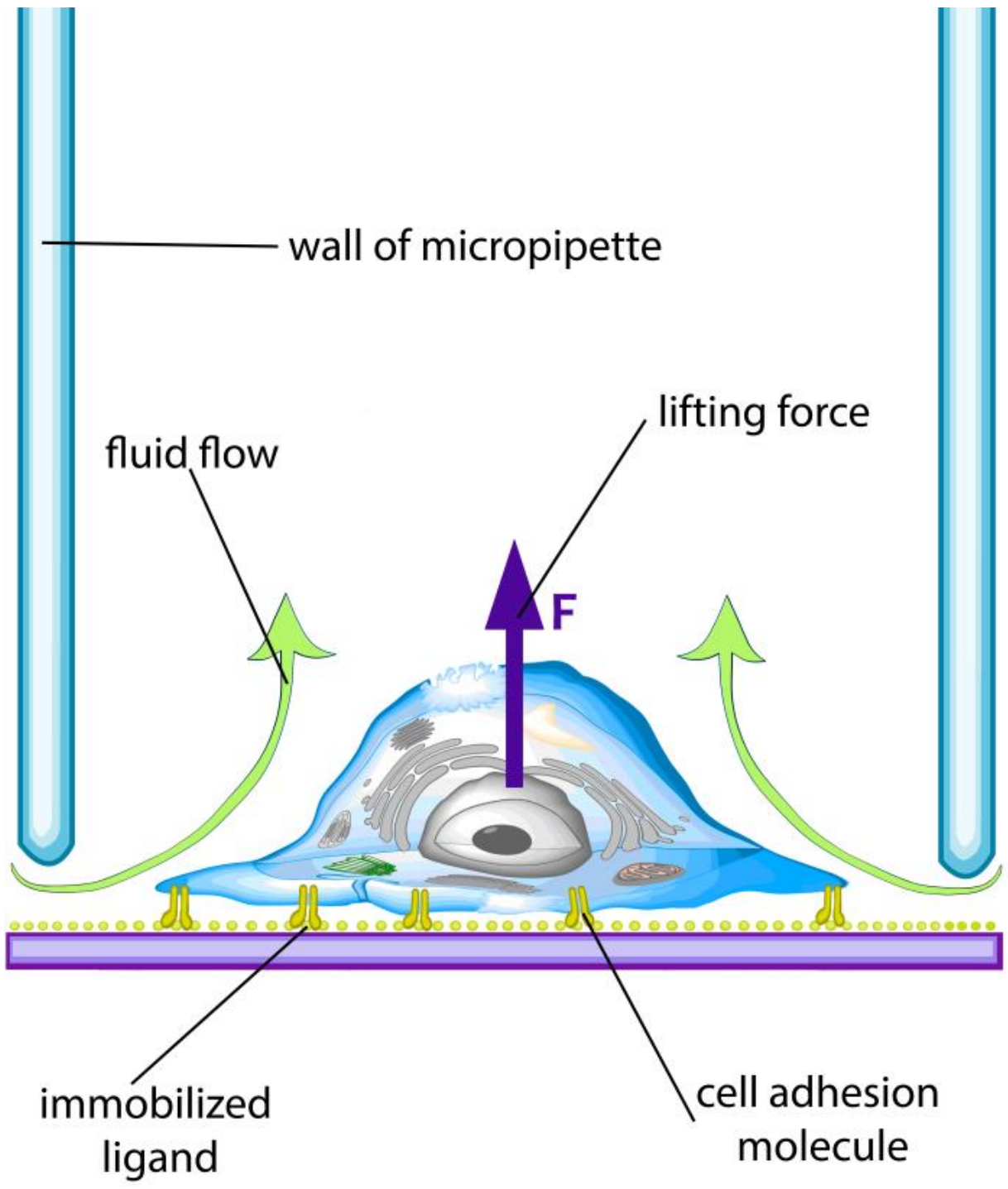

***Fig.11*** *Schematic illustration of the computer controlled micropipette. The cells are approached by a micropipette from above, then a negative pressure is applied generating an upward fluid flow. If the lifting force is greater than the adhesion force between the cell and the surface, the cell is aspired. By increasing the vacuum in steps, the adhesion force distribution can be measured on a relatively large population of cells as the computer guides the micropipette through the Petri dish. The bottom surface can be functionalized by specific natural or artificial ligands.*

### *a. Number of cells in an experiment*

Single cell measurements in a cell culture dish containing hundreds of cells can be performed. In our experiments the total number of human immune cells probed by the micropipette was 200-600 for each of the three cell types [23].

### *b. Typical force range*

Experimental vacuum value in the syringe can be converted to hydrodynamic lifting force acting on single cells on the basis of computer simulations of the flow in the micropipette. To estimate the lifting force acting on a real cell, the total force on a model cell (e.g., a hemisphere with a diameter of 20 μm) can be determined in 3D simulations.

In our experiments with human monocytes, adhesion force of most cells fell into the [0, 2] μN interval on the fibrinogen surface [23].

*c. Experiment duration*

Duration of the adhesion force measurements is typically 30 min [23].

*d. Advantages*

It can measure the adhesion force of individual cells with relatively high throughput: hundreds of cells in ~30 min, especially when compared to AFM or FluidFM. Measurements can be carried out on cells incubated on the specific surface for several hours or days to investigate physiological, potentially strong cell adhesion. The device can be mounted onto a normal inverted microscope. Both biologically and medically relevant results gained with the automated micropipette were reinforced in standard microfluidic shear stress channels. Automated micropipette offers a higher sensitivity than the measurement using the shear stress of a microfluidic channel [23] and it has less experimental side-effect than the shear stress channel. Cells usually become elongated and aligned to the direction of the flow in a shear stress channel. Similar effect was not observed in the experiments with the micropipette.

*e. Disadvantages*

A drawback of the technique as compared to AFM: to calculate the value of the adhesion force, hydrodynamic simulations depending on cell size (and less sensitively on cell shape) have to be carried out. However, computer simulations are not needed when different cell types or different treatments have to be compared without a need for a scaled value of the adhesion force.

*f. Main applications*

The adhesion force of cells attached to specific molecular surfaces can be accurately probed. Measurements can be carried out on cells incubated on the specific surface for several hours or days to investigate physiological cell adhesion. Cell-cell interactions can be also studied with this method [124].

## 3. Optical tweezers (OT)

Laser tweezers were developed for the microscopic manipulation of cells and organelles. Conventional manipulation techniques - including optical tweezers/optical traps, magnetic tweezers, acoustic traps and hydrodynamic flows - cannot achieve high sensitivity and high throughput at the same time.

Conventional OT (Table 3) (Fig.12) use a strongly focused Gaussian laser beam to trap and manipulate microscopic objects such as small dielectric spherical particles (bead) [127]. Trapping lasers operating in the near infrared regime (800-1100 nm) minimize optically induced damage in biological specimens [128]. High NA (typically 1.2-1.4 NA) microscopic objective lens is used to focus the trapping laser [127]. A small bead is captured in an optical trap. The bead is positioned to touch the surface of a cell, and then the bead is pulled away from the cell until the chemical bond breaks. The force between the bead and the cell is determined on the basis of the displacement of the bead from the focus perpendicular to the optical axis. Optical tweezers allow fine control of positioning (~10 nm for trap beam stability) and of forces (~ 0.1 pN resolution) on a wide range of particle sizes (25 nm to 25 µm) in a non-invasive manner [129],[130]. Owing to their precisely controlled force-exerting characteristics, OT are often used for a variety of mechanical force measurements in the pN range for single cells. Although most cells cannot be directly grabbed by the OT due to their size or shape, a small number of cell types such as yeast cells, RBCs and spermatozoa are readily tweezed and provide model systems for such force studies [131]. For those cells that cannot be directly tweezed, the use of microspheres as handles for force probes has allowed the measurement of cellular properties such as membrane tension [127].

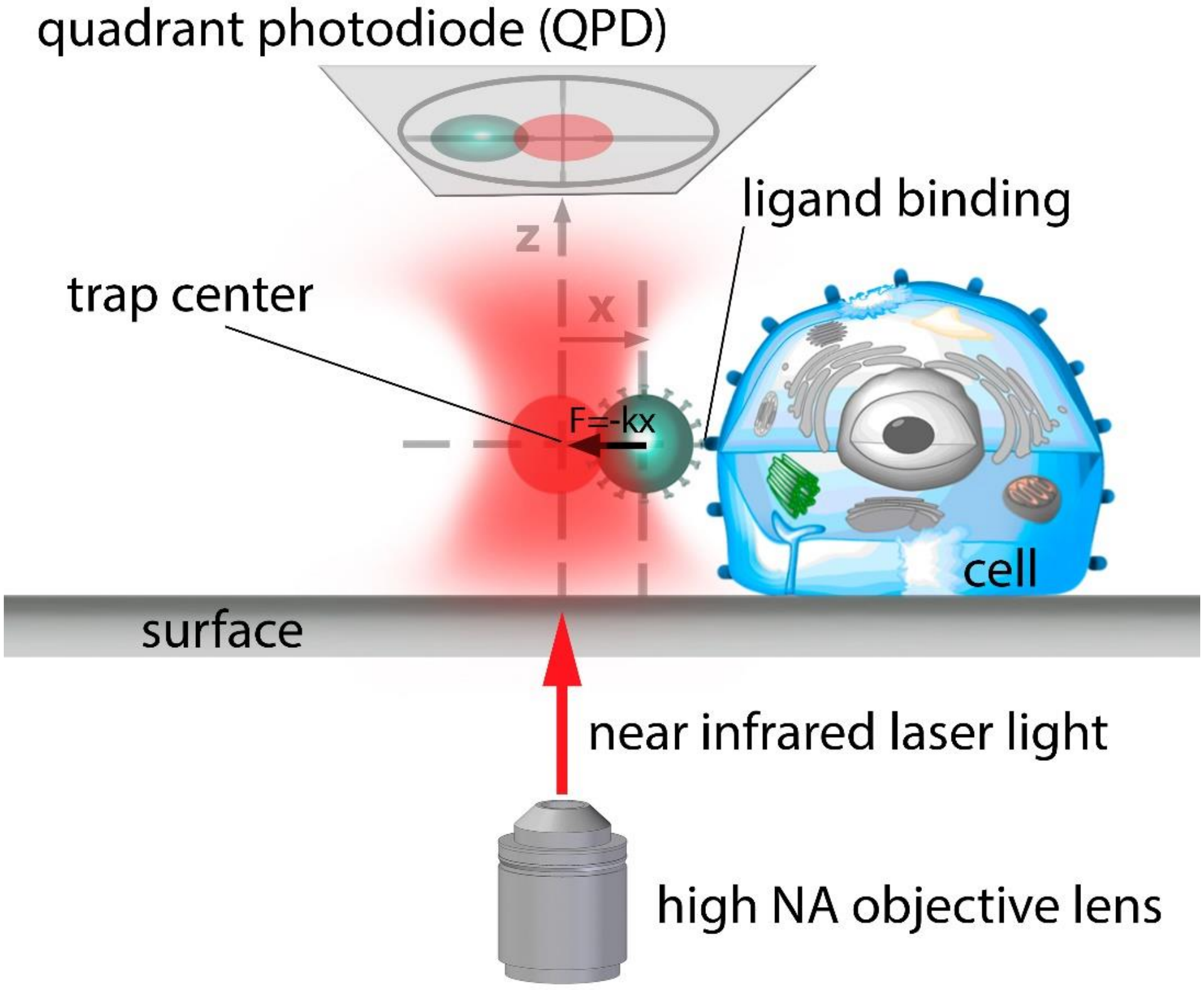

***Fig. 12*** *An optical trap is created by focusing a near infrared laser light to a diffraction-limited spot with a high numerical aperture (NA) microscope objective. The laser beam used for trapping has a 2-D Gaussian intensity distribution with an intensity gradient in the x- and y-plane (perpendicular to the beam axis). A particle (bead) is captured in the optical trap and positioned to touch the surface of a cell. Then the particle is moved away from the immobilized cell. The force between the bead and the cell is determined on the basis of the displacement (x) of the bead from the focus perpendicular to the optical axis. The particle displacement changes the intensity distribution of the transmitted infrared light. The shift of the intensity maximum can be detected with a quadrant photodiode (QPD). The signal of the QPD can be converted into units of displacement (nm) and force (pN).*

*a. Number of cells in an experiment*

1 cell/experiment.

*b. Typical force range*

0.1-100 pN [127]. The trapping force depends on the intensity of laser power, the shape of laser focus, the size and shape of the trapped particle and the index of refraction of the trapped particle relative to the surrounding medium. It is difficult to measure the trapping force directly but there are several ways to calibrate it [127]. The external forces applied to a single particle within an optical trap can push a single particle away from the focus, and the force from the OT draws the particle back to the center of the trap. In the equilibrium position, the external force equals the force from the OT. For a small displacement, the force from the OT, termed as restoring force, can be estimated by:

$$\mathbf{F} = -\mathbf{kX} \tag{6}$$

where $k$ is the trap stiffness and $X$ is the displacement of the particle from the center of the trap [127],[130].

*c. Experiment duration*

A bead carrying ligands on its surface is brought in contact with the cell. After an incubation time of a few seconds, it is pulled away from the cell applying stress to the chemical bonds [132]. When the cell itself is trapped for several minutes at high power (P > 300 mW), membrane stiffening can be observed, and the cell elasticity is affected. Thus the manipulations on the same cell should take normally less than 15 min [133].

*d. Advantages*

OT offer exceptionally high force sensitivity [106]. The technique can be applied to measure weak forces of single molecules that cannot be achieved by traditional AFM or most of the alternative tools [127]. High spatial resolution [130]. Non-contact force measurement [127],[131].

*e. Disadvantages*

Optical manipulation is limited in space due to focusing requirements [106]. Choice of the laser can be critical depending on the application. Laser absorption by the sample can lead to damage as the highly focused spot has an intensity in the range of $MW/cm^2$ [131]. Thus wavelength is a central parameter when cells are trapped: wavelengths below 800 nm can easily damage cells [134]. Thermal effect of the high laser intensity has to be considered [127],[135].

Trapping in cell extract or in a medium containing impurities is generally precluded as trapped impurities can distort or mask the position signal. Still the optical trapping of lipid vesicles within eukaryotic cells [136], and organelles in yeast cells [137] have been successfully implemented.
The applied force is limited to a maximum of 100 pN set by the maximum laser intensity in the specimen plane [130].

*f. Main applications*

Various cell types, including mammalian cells (red blood cells, nerve cells, gametes and stem cells), yeast cells [138],[139],[140] and bacteria, such as Escherichia coli have been studied [127].
In the binding experiments, the 3D manipulation capability is exploited to impose a specific interaction between the trapped object and a fixed partner, and to measure the force and displacement resulting from the interaction [130],[141]. A particle is decorated with ligand molecules binding to the cell surface. The optical trap pulls the particle away from the cell until the chemical bond breaks [132]. Measurement of the force and displacement of optically trapped kinesin coated beads moving along fixed microtubules was pioneered by Block *et al.* [142],[143]. The binding probability and unbinding force was measured between virus coated beads and erythrocytes. The binding force of single fibrinogen/fibronectin-integrin pairs could also be quantified in living cells [143],[144]. OT can capture the force characteristics of intermolecular bonds on the cell surface [132].
In the so-called tether-pulling experiments, OT have been applied to study the mechanical properties of cell membranes. Tether extraction is an accurate method to quantitatively characterize the plasma membrane. To form a membrane tether, micrometer-sized particles (beads) are used to grab the cell membrane. A bead trapped by the OT is held on the cell plasma membrane for a few seconds, and then moved away from the cell to pull out a membrane tether, a thin cylindrical strand of plasma membrane between the bead and the cell [145],[146]. The force needed to manipulate the bead can be measured by OT [127]. Thus the bending rigidity of the membrane can be determined [132].

OT can probe the viscoelastic properties of whole cells under physiological condition. For this two optical traps are applied to attach beads to two opposite points of the cell. It can give an insight into the internal structure and organization of the cell [132].

# 4. Atomic force and fluidic force microscopy

Atomic force microscopy (AFM) developed by Binnig et al. (1986) [147] can be applied to capture single cells with its functionalized cantilever gently pressed to the cell (Fig.13 A). This in turn converts the living cell to a probe brought into contact with a functionalized surface or another cell (Fig. 13 C). Subsequently, the cantilever is pulled back at a constant speed, detaching the cells from its binding side. Deflection of the cantilever is proportional to the force acting between the cell and the substrate. It is recorded as a force-distance curve [43]. Deflection of the cantilever is detected with a laser beam focused on the top side of the cantilever and reflected into a quadrant photodiode [148].

In the force spectroscopy mode, AFM (Table 4) acts as a versatile tool to probe nanomechanical properties and to extract quantitative parameters of biological systems, including tissues, cells, proteins and nucleic acids, and of biomimetic systems, such as functionalized surfaces or matrices. AFM-based force spectroscopy [43],[149] involves an atomic force microscope tip to locally record interactions with the sample or its nanomechanical properties. Approach and retraction force–distance curves characterize the sample deformation and tip–sample adhesion, respectively [150]. Analysis of the approach force–distance curve, and in particular the region describing indentation, allows properties including deformation, elasticity and dissipation to be determined. The retraction curve quantifies the adhesion force between the tip and sample.

Single- molecule force spectroscopy (SMFS) [151],[152] is frequently applied to detect the binding strength of ligand receptor pairs. It was applied to measure the force required to unbind streptavidin and biotin [153], it was quickly recognized that measuring rupture forces provides information about the kinetic properties of a bond.

Single-cell force spectroscopy (SCFS) [154] measures the adhesion of a single cell to a biointerface, which can be tissue, another cell or a surface functionalized with ligands. A single cell is attached to the free end of a tipless cantilever [150]. A number of methods have been developed to attach the cell onto the AFM cantilever and to allow probing cell-cell or cell-substrate interactions [148]. Specific receptor-ligand interactions [155],[156] electrostatic interactions [157], glue [158] or chemical fixation [159] can all be applied. An important issue when applying these protocols is to make sure that the native surface of cells is not altered or denatured. In this respect, an appropriate approach is to attach individual cells to the AFM cantilever with lectin. This method allowed the measurement of the adhesion force between two neighboring cells of D. discoideum on the single-molecular level [148],[155]. After that, the functionalized cantilever is lowered into contact with a trypsinized cell, which readily attaches to the cantilever. Then, the probe cell is brought into contact with a biointerface for a given contact time and force, and then withdrawn while a force–distance curve is recorded. Analysis of the force curve provides the maximum adhesion force of the cell. This approach allows investigate how cells strengthen adhesion to ECM proteins or other substrates [150].

AFM-based force spectroscopic modes enable the characterization of single receptor–ligand bonds, protein unfolding and refolding, and the mechanoelastic properties of peptides, nucleic acids, sugars and polymers. Cell adhesion to the interface of substrates, other cells or tissues can also be quantified using such modes.

AFM can be used in 2 ways: it can either measure the adhesion of the cantilever-attached cell to the substrate surface or the adhesion of the cell immobilized on the substrate to the cantilever [149] .

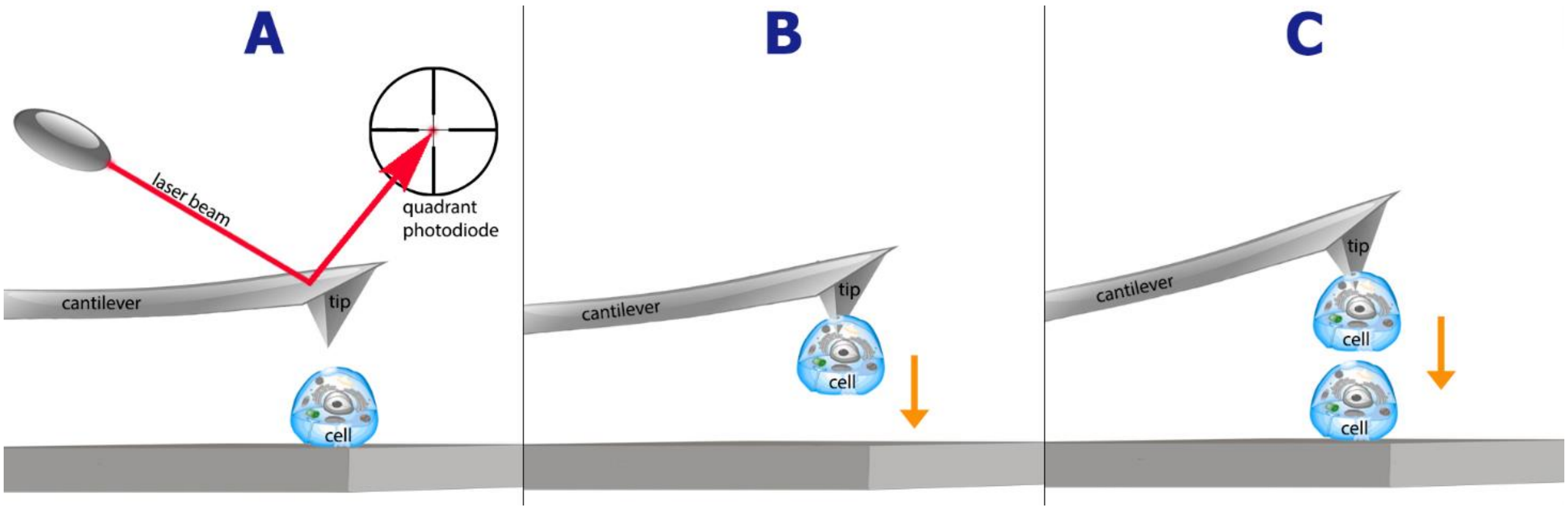

***Fig.13*** *Schematic illustration of atomic force microscopy applied for cell adhesion measurements. The interaction between the cell and the tip is measured through the displacement of the laser beam on the photodiode caused by the deflection of the cantilever. The latter acts as a spring probing the force acting between the tip and the cell. The tip surface can be functionalized with a molecule of interest then it is pushed to the cell for a predetermined time that allows the receptor-ligand bonds to form (**panel A**). As the tip is pulled back with a controllable speed, the distance-force curve can be registered by measuring the signal on the photodiodes. As an alternative to tip functionalization with biomolecules, a living cell can be attached to the tip (**panel B**). Then the cell attached to the cantilever is pushed to the functionalized substrate surface or to the other cell on the substrate, thus allowing the direct measurement of cell-cell adhesion (**panel C**). Since both the spatial resolution and force sensitivity are high, the AFM was the first method capable of measuring cell adhesion on the single receptor level.*

## 4.1 AFM to probe force normal to the surface

*a. Number of cells in an experiment*

Each cell requires a separate cantilever that must be functionalized (typically ~30 min) and calibrated, which impedes the ability to obtain high-throughput measurements [48]. At least three force curves have to be recorded at different locations in each cell. Although it is beneficial to take multiple curves on each cell for reliable statistical data, taking too many force curves can lead to changes in cell stiffness due to the mechanical stress caused by the AFM probe. AFM yields only a few measured cells (from 1 to 5) per condition in most studies [160].

*b. Typical force range*

The AFM (Binnig et al., 1986) can be used to measure nano-newton to pico-newton forces and micrometer to 0.1 nm displacements [161]. Imaging in aqueous solution allows observation of biomolecules under physiological conditions and estimate strong, capillary forces between the sample and probe when the imaging is done in air [162],[163]. In order to prevent deformation of weak biological sample, a vertical force is used (50-100 pN=minimum force). At 100-1000 pN force, the biological sample is reversibly deformed, and above this value irreversibly [163],[164],[165]. AFM can be used to analyze accurately and then remove genetic material from the chromosome [162],[166]. Antibodies that specifically bind to the cytoplasmatic membrane surface can be removed with smaller than 0.8 nN [162],[163],[167]. In the future it is expected that the use of smaller cantilever improves the resolution, allowing measurement of the smaller non-binding forces.

AFM-based SCFS (single cell force spectroscopy) enables measurement of wider force range (5 pN-100 nN) [149].

An individual WM115 cell was bound to a lectin functionalized cantilever with a given force (500 pN) by Puech et al. and then the cantilever with the bound cell was slowly retracted 100 μm from the substrate. This allowed a strong adhesion of the cell to the cantilever (5-15-min). After that, adhesion of an individual, captured melanoma cell and a model endothelial layer of HUVEC cells were tested. A given force (few 100 pN) was applied on the cell layer for a certain time (range of sec). Finally, probe cell was separated from the surface while detachment event was recorded [44]. This work described that, AFM is a sensitive and quantitative method to measure long-distance cell-adhesion forces.

A D. discoideum cell was bound to the AFM cantilever by Benoit et al [155]. Then a target cell was placed at the bottom of the Petri dish and it was approached while a defined repulsive contact force was established. This force was held constant to enable establishment of cell adhesion. While the cantilever was retracted, it was recorded the force as a function of the distance until connection between the cells was broken. It was allowed the interaction for 20 s at the force of 150 pN. Adhesion between these cells produced unbinding forces of the order of 1 nN [155].

SCFS (single-cell force spectroscopy) was used by Taubenberger et al. [168] to investigate the α2β1-mediated cell adhesion to collagen I.; they compared adhesion of CHO-WT and CHO-A2 cells to collagen. It was found that, CHO-A2 cells adhered to collagen (40-600 pN) strongly, than another cell (<50 pN). When contact time was < 120 s, usually 2-5 force cycles/cell were carried out, whereas for longer contact time, 1-2 force curves/cell were recorded [168].

Lehenkari and Horton determined integrin-ligand interaction of osteoclasts. Integrin-binding forces were measured in intact cells by AFM for several RGD-containing ligands with a range of 32-97 pN. It was reported that, AFM can be applied in cell biology studies and it provides an opportunity to analyses receptor-ligand interactions in the cell membrane [10].

*c. Experiment duration*

First, a chemical functionalization of the cantilever is needed allowing irreversibly ‘‘glue’’ the cell of interest to the cantilever; where the lever is coated with wheatgerm agglutinin for at least 1 h. Consequently, each cell requires a separate cantilever that must be functionalized (typically ~30 min) and calibrated, which impedes the ability to obtain high-throughput measurements [48]. The AFM cantilever is positioned using a piezoelectric crystal, and its deflection is measured by laser reflection onto a split photodiode. Positioning precision in the z-direction is 1 Å, and force sensitivity is within 3 pN. The cantilever is moved with a velocity of $2.5 \pm 0.5$ $\mu m\ s^{-1}$ [155].

Cells are immobilized on a substrate or on the force sensor itself [155] (Fig.13 A,B). The sensor surface is functionalized with an adhesive molecule [169] to attach the cell to it. Cells immobilized on the sensor were used to investigate cell-cell interaction [161] (Fig. 13 C). To avoid scattering of the laser beam of the detection system, non- or weakly-adherent cells are removed by gently rinsing the dish. An attached cell is slightly loosened by pushing its flank with the side of the cantilever. The extreme end of the lever is then lowered onto the cell at a force of a few nN and hold in contact for ~ 30 s to allow the lectin on the lever to bind; the cell is then lifted off the bottom of the dish [155].

Using AFM, it needs ~10 min for proper cell–cantilever interaction. In this technique, the force range is limited by the force with which the cell is bound to the cantilever, which usually results in the application of only relatively short contact times (msec-20min) [43],[170] between the cell and the substrate of less than one hour before the adhesion force exceeds the detectable range [48],[161].

*d. Advantages*

AFM can measure adhesion from a single molecular event in intact cells under physiological conditions. Both spatial and force resolutions are high. AFM has force sensitivity in the pN range and nm positioning accuracy, therefore atomic force microscopy is a powerful device to explore dynamics and strength of interactions between individual ligands and receptors. These studies require adhesion of specific biomolecules or cells onto the AFM tip or to a solid surface. Currently, AFM is the only force- measurement technique which can map and analyze individual receptors with nanoscale lateral resolution [148]. AFM has several unique properties: it can be operated in solution, allowing observation of biological structures in native environment [171]; individual proteins can be observed at a resolution >1 nm [165],[172] conformational changes of single biomolecules can be directly monitored [162],[165],[170],[173].

*e. Disadvantages*

State- of- the- art AFM measurements have limitations, for example adhesion measurements, which use single cells are time-consuming, costly methods. Only one cell can be characterized at a time. Each cell requires a separate cantilever which must be calibrated and functionalized [43],[48],[149] impeding to obtain high-throughput measurements [170]. Chemical attachment of the cell to the cantilever is cumbersome (~30 min) and it can alter the physiology of the cell [161],[174]. High number of detachment force-distance curves need to be measured to gain reliable statistics which restricts the length of the contact time. Usually only a relatively short contact time is established between the cell and the substrate (msec-20min) [43],[170]. Another challenge is to develop a fast scanning AFM as the temporal resolution of the method is limited [148]. Thermal drift is almost unavoidable in AFM experiments which complicates long-contact-time experiments (>20min). Strong adhesion of the cells falls out of the force range of the technique after a longer contact time (>1h).

*f. Main applications*

Due to its wide force detection range, it can measure both the adhesion of a whole cell and interaction [175] in a ligand-receptor system [43],[149]. It can study the dynamic formation of cellular adhesion. It has been adapted to measure cell-cell and cell-substrate adhesions. AFM can be combined with modern optical imaging techniques [43]. It can quantify forces guiding microbial cell adhesion [78],[175],[176],[177],[178].

## 4.2 AFM to probe tangential/lateral force

Manipulation force microscopy [2] (Table 4), an atomic force microscopy can measure the force that dislodges micro-sized objects attached to the surface. The technique applies laser beam deflection and an inclined AFM cantilever to measure the force. The cell is brought into contact with the cantilever, and a tangential force is applied gradually. Finally, the cell is released from the surface as all adhesion bonds are broken.

*a. Number of cells in an experiment*

Adhesion strength of cervical carcinoma cells were examined by this technique. In an experiment, force measurement on ~200-300 cells were carried out [2].

*b. Typical force range*

Adhesion force of several hundred nN was measured on different surfaces [2]. A force of 19 nN was required to detach E. faecalis cells from hydrophobic materials, and 6 nN to detach from hydrophilic materials [176]. Chang and Hammer in their numerical simulations of functionalized microbeads found that the lateral detachment force is several times lower than the perpendicular detachment force. Thus a tangential force is considered to be more effective to detach a particle than normal forces [177]. The reason is thought to be the low ratio of the bond length to the bead radius.

*c. Experiment duration*

Adhesion time is usually between 10 and 90 min [2].

*d. Advantages*

Studies have shown that the lateral force required to displace a bacterial cell is considerably (up to 10 times) smaller than the perpendicular force [177],[178]. Enables the real-time imaging of bacterial adhesion and aggregation under physiologic conditions or affected by antibiotics [176].

*e. Disadvantages*

Interpretation of the experimental data is not straightforward [2]. The measured force depends strongly on the details of the experiment: the shape of the tip, the scanning speed, the torsional spring constant of the cantilever, the exact direction of the applied force, and the nature of the binding between the cell (or bead) and the substrate [78],[179].

*f. Main applications*

Binding force of protein-covered silica spheres adsorbed to polystyrene surfaces [180]. Adhesion force of individual cervix cells to various substrates [2]. Characterization of medical-grade polymers and their resistance to microbial adhesion and biofilm formation [176].

### 4.3 Fluidic force microscopy (FluidFM)

FluidFM (fluidic force microscope) is a unique micromechanical and microfluidic device that combines the force-controlled high spatial precision of AFM with the capability of direct liquid delivery by microfluidics. The needle-like AFM probe or the bare cantilever of the FluidFM contains a fluidic microchannel inside (Fig.14). It is directly connected to an external fluid circuit controlling the pressure and thus the flow in the microchannel. This novel technique can both mechanically manipulate living cells and microinject biochemical reagents into them [181]. The whole FluidFM BOT system involved an inverted optical microscope [182]. Individual microbial (S.cerevisiae, C.albicans) and mammalian cells (HEK, HeLa) were used to measure the cell adhesion force [48]. FluidFM (Table 4) reuse the same probe for the measurement of multiple cells to record single cell force spectroscopy (SCFS) curves reducing the time required to obtain statistically relevant data compared with conventional AFM [48]. Trapping the cell to the cantilever occurs within a few seconds before the SCFS measurement. Then the cells can interact with the substrate or the other cell for a longer time.

During the measurement, first the cells are selected, and then it is approached with a set point (10 nN for yeast cells and 50 nN for mammalian cells), after that it is followed a pause (5 s for yeast cell and 3 s for mammalian cell) with force feedback using underpressure and the cell is immobilized to the cantilever. While using AFM, it requires ~10 min for cell-cantilever interaction but with FluidFM needs only 5 sec underpressure for cell-cantilever contact. In the conventional AFM, cells are immobilized to the cantilever, while FluidFM to fix the cell to the cantilever, apply underpressure. Here chemical fixation of the cell to the cantilever is replaced so that cells are sucked into the aperture of a hollow cantilever (few sec). The cell release from the substrate so that it is applied an overpressure pulse makes the immediate reutilization of the cantilever possible, thereby resulting in the ability to perform serial measurements. Potthoff et al. have shown that FluidFM can increase the amount of data that can be recorded in a single day. Compared to the conventional SCFS, FluidFM based SCFS can be performed up to tenfold as many experiments / day and 200 experiments can be carried out with this technique compared to the conventional cantilevers [99]. This approach is similar to the microinjection with glass pipette, but there are differences. Microinjection requires optical microscopy. Cells are often damaged, however using AFM precise force feedback reduces potential damage to the cell [181].

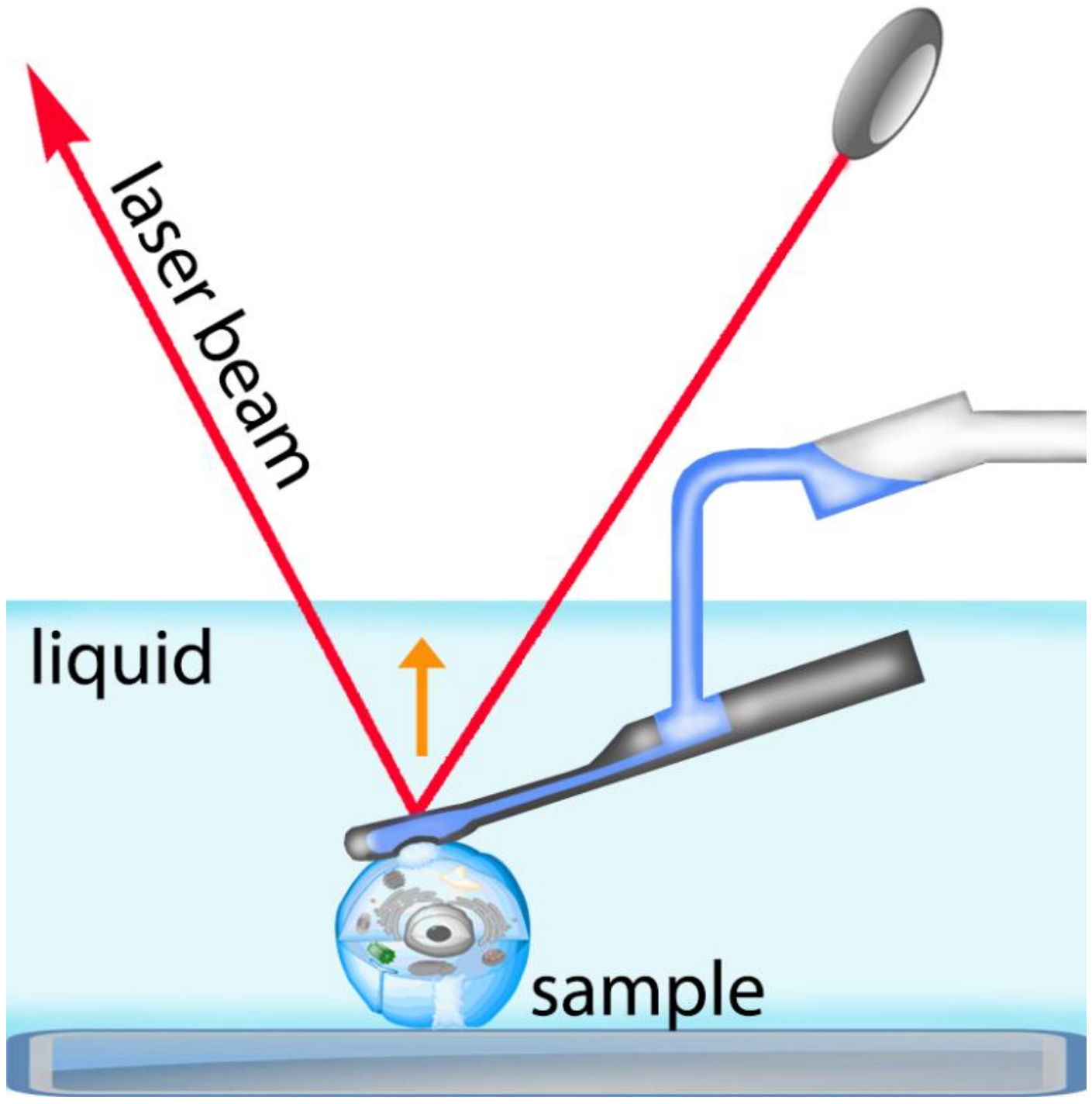

***Fig.14*** *Schematic illustration of fluidic force microscopy. The nanofluidic channel inside the cantilever allows the grabbing of the cell by the cantilever or tip using a negative pressure. When the upper side of the cell is immobilized this way, the tip is moved upwards picking the cell up from the surface. This method can be used to directly measure the complete adhesion force and energy between the adhered cell and the surface in a less difficult way than using the traditional AFM method. Another advantage is the rapidity of the measurements, as the tip can be reused after a ~ 5 min washing step. The upper limit of the measured force is set by the aperture of the fluidic microchannel and the spring constant of the cantilever. Since there are multiple tip designs available, the same device can be used for other applications, such as cell injection, lithography, etc.*

*a. Number of cells in an experiment*

It was compared the maximum adhesion strength of 2 types of yeast cells (S.cerevisiae, C. albicans) to hydrophobic and hydrophilic surface, after an adhesion time of 15 min. Then HeLa (n=11-12 cells were examined serially) and HEK cells (n=8-9cells) were compared on glass and fibronectin-coated surfaces. Up to 200 yeast and 20 mammalian cells per probe can be performed with this technique [48]. In half an hour, ~10 cells can be studied using this technique.

*b. Typical force range*

The measured adhesion forces were between 500 pN-1.6 µN [48].

It has been shown that, adhesion force is increased as a function of incubation time and dependent on the temperature. It has been demonstrated rapid and serial yeast cell adhesion which can available with FluidFM so it can be investigated adhesion strength to the specific substrates and it can be determined long-term adhesion interactions.

It was studied the adhesion of C. albicans to hydrophobic surface that was 39 nN and to hydrophilic surface was 10 nN. Adhesion strength of S.cerevisiae hydrophilic and hydrophobic surfaces was 2 and 5 nN, respectively. In both cases the same behavior was occurred; adhesion to the hydrophobic surface was stronger.

It was studied whether FluidFM is suitable to detach mammalian cells from standard surfaces while long-contact time is established between cell and substrate but only a minimum time is required to fix the cell to the cantilever. HeLa cells on fibronectin surface showed 40-fold greater adhesion strength as compared to yeast cells. HeLa was compared to another cell line, HEK, and adhesion of this cell to fibronectin was 53 nN which is much smaller than in case of HeLa [48].

*c. Experiment duration*

Compared with the conventional SCFS, using FluidFM-based SCFS allows to carry out 10-times more experiments (using FluidFM can be performed up to 200 attempts while using AFM only 20 attempts). Only a few minutes is needed to target, immobilize, and release the cell as well as to change the cantilever position to the next cell [48].

*d. Advantages*

It can be measured the adhesion force directly. It has been demonstrated a fast and serial yeast cell adhesion measurements. It can be monitored adhesion of single cells to specific substrates and can be quantified the long-term adhesion interactions. The technique enables long contact times (orders of magnitude) between the cells and substrates. It has been demonstrated the universality and broad applicability of this method for different cell types. It can be measured higher cell-substrate adhesion forces (one order of magnitude), than using AFM [43]. Much shorter measurement time and many more adhesion measurements can be performed with it. Instead of chemically immobilizing the cells to the cantilever as in conventional AFM experiments, it is applied underpressure to fix the cell to the cantilever aperture [183].

*e. Disadvantages*

Using mammalian cells, it was unable to obtain the same quantify of serial measurements which were monitored with yeast cells, but it could be recorded 10 times more force curves than with conventional SCFS [48]. Throughput is lower [181] than that of the automated micropipette. Microfabricated cantilevers come with a high measurement cost. Cells come into direct contact with the cantilever potentially perturbing or damaging cells [181].

*f. Main applications*

The universality and versatility of the FluidFM opens the way for original experiments in physics, materials, sciences, chemistry and molecular electronics. The method reduces the time required to obtain statistically relevant data compared with conventional SCFS. The use of FluidFM based SCFS enables AFM based adhesion force measurements [48],[184]. This is a universally applicable method for living cells. It can measure the adhesion force of those cells that were difficult or impossible to measure with conventional AFM [48]. Microinjection of cells [185]. It can be used for printing of 2D surface patterns with various formations [186] or for precisely controlled template-free 3D micro- and nanoprinting [187],[188]. Micropatterning of living mammalian cells on carboxymethyl-dextran (CMD) hydrogel layers was presented using the FluidFM BOT technology [182].

# 5. Cell traction force microscopy and FRET force sensors

## 5.1 Cell traction force microscopy

**Cell traction force microscopy (CTFM)** [49] (Table 5) enables the force (mechanical stress) field generated by single cells on an elastic substrate to be reconstructed with a resolution of ~1-10 μm. The workflow of the technique is built up of four subsequent procedures, namely: 1) preparation and characterization of the substrate; 2) microscopic imaging of cells and deformation markers inside the elastic substrate; 3) computational recovery of the displacement field of the substrate; 4) computational determination of the traction stress field from the displacement field, based on an elastic model.

Most often, the substrate is a polyacrilamide (PA) gel coated with proteins that specifically promote cell adhesion (Fig. 15). PA has the advantage that it is linearly elastic within a wide range of mechanical stresses and thus completely recovers after stress removal [189]. Furthermore, by varying the degree of crosslinking in the PA gel, its rigidity can straightforwardly be tuned in the physiologically relevant range of 0.1-100 kPa ($1kPa = 1nN\mu m^{-2}$). The PA gel is embedded with randomly distributed fluorescent beads (usually 200-500 nm in size) which are then used as markers of the substrate deformation.

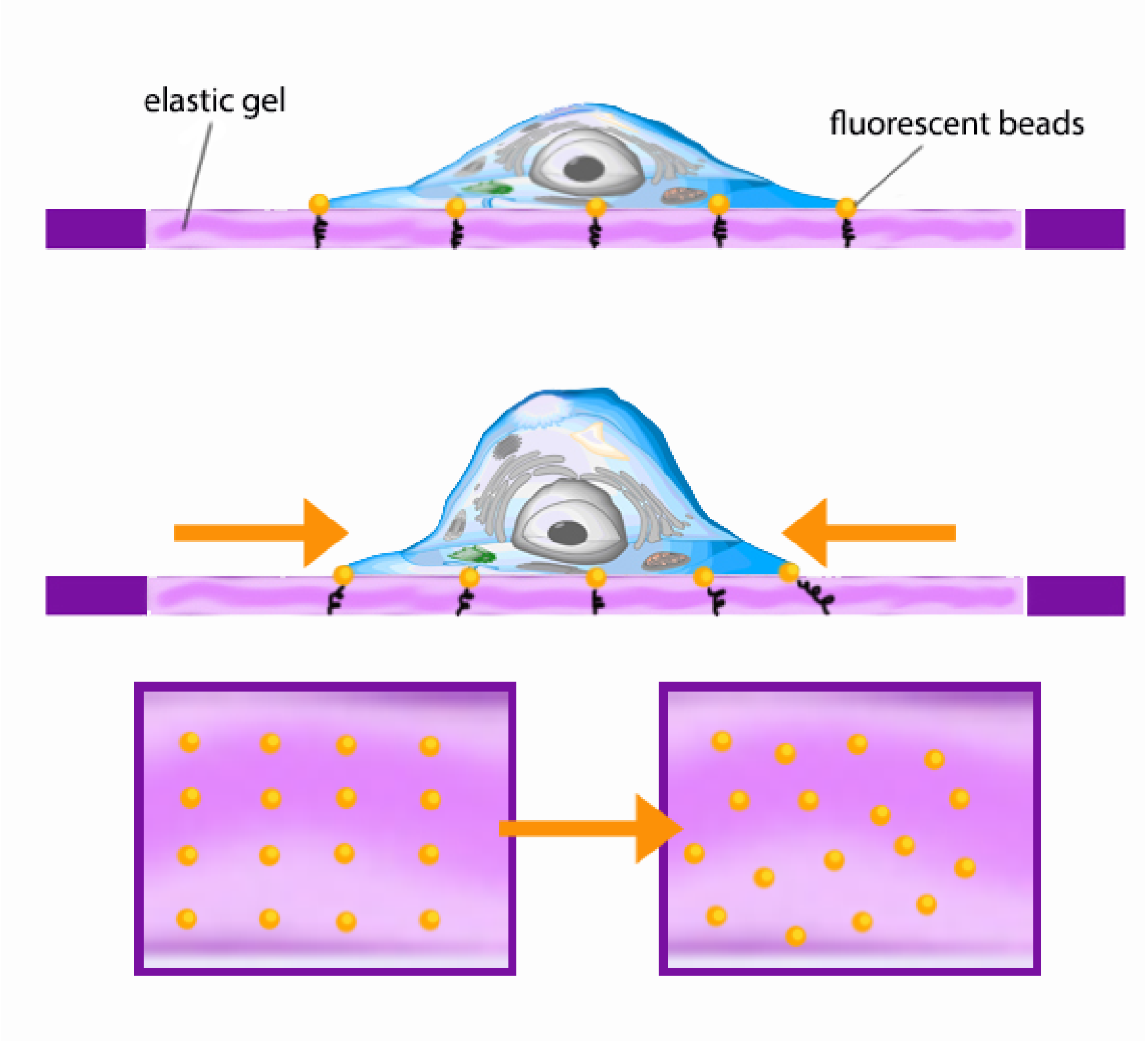

***Fig.15*** *Schematic illustration of the principle of cell traction force microscopy. The substrate (in purple) is randomly embedded with fluorescent beads with a diameter of around 500 nm. Then the cells are placed on the surface and let to adhere for a predetermined time. As the cells interact with the adhesion proteins on the polyacrilamide substrate, a force is exerted that rearranges the previously recorded bead distribution. The fluorescent patterns before and after cell adhesion are captured by wide field microscopy. From this displacement field, the traction field can be reconstructed by an elastic computational model with a spatial resolution of ~1 μm. Using the 3D version of the method, the traction field can be measured in 3D cell cultures. CTFM has been successfully applied for the mapping of individual cell's traction fields as well as for the examination of the migration of cell groups.*

### *b. Typical force range*

Traction force microscopy relies on culturing cells on soft compliant polymer films that deform under tension [190]. These deformations are then deconstructed using computational finite element analysis to calculate the lateral force vectors applied to micrometer elements. Using this method, the literature estimate for integrin tension within focal adhesions is approximately 2-3 pN per receptor [191]. The typical force range is nN in magnitude [191].

*c. Experiment duration*

Up to several hours when combined with time-lapse imaging. After pre-incubating the cells on the substrate for an arbitrary time, comes the imaging stage of the workflow when at least a set of three photomicrographs are taken. First the cell-substrate contact is imaged, often in simple phase-contrast mode. Next, the microbead distribution near the surface of the substrate is captured in fluorescent mode for the cases when i) the cells exert forces onto the substrate (force-loaded image) and when ii) they have been removed from it via trypsinization (force-free image) [192],[193]. Fluorescent imaging is generally done with a wide field epifluorescent microscope.

*d. Advantages*

Since its introduction in 1996, CTFM has undergone impressive developments. Innovations at the substrate level gave rise to CTFM variants which characterize the substrate deformations in a more or less different manner as described above [194]. For instance, a CTFM variant utilizes a PDMS substrate with a flexible array of single micropillars which act as independent tiny strings. Unlike the original CTFM, this variant has the advantage that traction forces can be easily calculated from the deflections of the micropillars [194]. On the other hand, cell adhesion is artificially constrained to discrete islands, and the spatial resolution of the mapping of traction forces is limited by the number of the micropillars. Recent technological and/or computational advancements have led to the development of i) high resolution CTFM offering a spatial resolution of ~1 μm [190] or even super-resolution [195]; ii) 2.5D CTFM, which enables all components (x,y,z) of the traction generated by a cell on a 2D substrate to be determined [192]; and iii) 3D CTFM, which permits traction force measurements for the case of cells embedded in a 3D matrix [196],[197].

*e. Disadvantages*

In order to recover the displacement field, the pixels of the force-loaded and force-free images have to be one by one appropriately matched (image registration problem). The most straightforward method would be to directly monitor the movements of the microbeads but given the high number (few thousands) of microbeads, this task is computationally demanding and is thus rarely used. Albeit alternatives to this method have been proposed, image registration methods continue to share some common problems, such as the deterioration of the displacement field when the beads become more sparsely distributed, or the difficulty of error estimation (for reviews see [193] and [194] ).

Nevertheless, once the displacement field is calculated, the deconvolution of the traction field remains the last step in CTFM to be carried out. It is critical to deliberately choose an elastic model to formulate the relation between the displacement and traction fields. Generally, the substrate is considered as a semi-infinite half space, thus only the substrate surface enters analysis and the simple Boussinesq equation [198] has to be solved . However, it has not got a unique solution and therefore regularization terms have to be added to exclude non-physical solutions. Furthermore, the inversion of the elastic equations, and thus the computed force field are very sensitive to noises. To circumvent this disadvantage, more sophisticated approaches have been proposed [193],[194].

*f. Main applications*

CTFM has hitherto been applied to map the traction field of individual cells [199],[200] and that of cell aggregates [201]; to detect phenotypic changes that are accompanied by a change in traction forces during cell differentiation [202]; and to investigate the migration of individual cells [203], as well as the collective migration of a sheet of cells [189],[204].

### 5.2 FRET force sensors

FRET (Förster/Fluorescence resonance energy transfer) (Table 5) is a form of non-radiative energy transfer between two fluorophores (donor and acceptor) [205] due to dipole-dipole interaction. Fluorescent organic molecules have been widely used as donors and acceptors and they offer advantages such as small size, compatibility with numerous and simple covalent coupling strategies, and a relatively strong optical signal. Careful selection of an appropriate donor–acceptor pair ensures high transfer efficiency and provides two measurable parameters: quenched donor photoemission and enhanced acceptor fluorescence [206].

FRET probability depends non-linearly on the distance between the donor and acceptor. This technique, capable of measuring distances on the 2-8 nm scale, relies on the distance-dependent energy transfer between the donor and acceptor fluorophores [207]. It is widely applied for mapping large scale protein structures as a ‘‘molecular ruler’’[205],[208],[209]. FRET measurements can follow receptor–ligand interactions, changes in protein conformation upon binding a target analyte or can be utilized to reveal the response to changes in the solution conditions (e.g., temperature or pH). It can also monitor the nanometer scale displacements of cell adhesion ligands.

A molecular force sensor can be constructed by attaching the donor and acceptor to the ends of a molecular spring: a flexible chain of atoms, usually a polymer. The distance of the donor and acceptor changes, when the molecular spring is under tension [50], and thus the tension can be read out from the FRET signal (Fig.16). Tension across the cellular receptor of interest leads to an elongation of the force-sensitive unit, which then can be microscopically detected [210]. FRET-based biosensors need high sensitivity microscopy and appropriate data analysis algorithms to determine the force in cells [211]. The most frequently used method is based on intensity measurements, in which the donor fluorophore is excited and the emission intensities of donor and acceptor fluorophore are used to calculate the FRET ratio. However, these measurements do not readily yield quantitative information on FRET efficiency, they are sensitive to the experimental settings and require careful image analysis. Alternatively, fluorescence lifetime imaging microscopy (FLIM) can be used to calculate FRET efficiencies from the donor lifetime in the presence or absence of the acceptor [211].

The molecular tension sensors can be divided into two categories, those that are genetically engineered and expressed within living cells (genetically encoded molecular tension sensors, GETS) and those that are anchored to a surface to probe cellular receptor forces at the interface between living cells and their external ligands (immobilized tension sensors) [212].

*a. Number of cells in an experiment*

Analysis was performed by using measurements from a minimum of 10 cells at each condition [209].

*b. Typical force range*

As FRET force sensors measure the tension built up in single molecules, the overall adhesion force of the cell cannot be simply determined on the basis of FRET microscopy. FRET-based tension sensors provide piconewton (pN) sensitivity within cells [210]. GETS are limited to the detection of forces within the range of 1 to 7 pN [212]. In 2010, Grashoff et al. designed a tension sensor module (TSMod) resolving forces lower than 6 pN. The talin, a molecule that connects integrin receptors with the actin cytoskeleton through multiple interaction sites experiences mechanical forces of ~ 7–10 pN [210]. The tension across vinculin in stationary focal adhesions (FAs) was measured to be ~ 2.5 pN [50],[212]. In the last years, TSMod-s gained popularity [50] and have been used to determine tension across the adhesion protein vinculin [50],[213], cadherins [214],[215],[216], PECAM-1 [216], spectrin [217] or the glycoprotein MUC-1 [217].

Immobilization of molecular tension sensors to a solid support allows forces between cell membrane receptors and their extracellular ligands to be investigated. Integrin receptors exert an adhesion force of 1-5 pN to their ligands [218].

*c. Experiment duration*

~ 20 min [209].

*d. Advantages*

Nanoscale interactions can be detected on the basis of the FRET [219] signal. FRET can be applied as a molecular force sensor built into single biomolecules. Genetic insertion of such a tension sensor module into the protein of interest and the expression of the resulting construct in cells allows the analysis of molecular forces in living cells [211]. It allows the measurement of single molecule reaction trajectories from about 1 millisecond to several minutes [220], [221].

FRET does not require mechanical perturbation of cells, which could alter cellular traction and adhesion forces. FRET can study how cells manipulate the ligands to which they adhere, and simultaneously determine cellular traction forces without perturbing the adhesion events [209].

A fundamental advantage of the technique is to measure the internal distance in the molecular frame rather than in the laboratory frame and hence it is largely immune to instrumental noise and drift [220].

The technique is adaptable to a wide variety of instrumentations, including fluorescence spectroscopes, conventional, total internal reflection (TIRF), confocal microscopes, and FACS [219].

*e. Disadvantages*

Artificial molecular force sensors are needed to be constructed and built into the cellular system, in most cases into a specific protein. Practically only one (or very few) protein type(s) can be measured in an experiment. Thus the overall cellular traction or adhesion force exerted by a large number of different proteins cannot be determined. The selection of an appropriate elastic molecular element of the force sensor is critical. The elastic linker has to be short or the increase in linker length has to be sufficiently large so that the applied tension can be observed as a change in FRET efficiency [211],[212]. FRET requires spectrally matched fluorophores. An ideal fluorophore for single molecule studies must be bright, photostable, small and water soluble. A problem inherent to any molecular force sensor is that its insertion can interfere with the properties of the host protein, and the host protein can interfere with the function of the force sensor [50].

*f. Main applications*

FRET can be used as a molecular ruler to monitor the nanometer scale displacements between adhesion ligands, and the corresponding traction force between integrin receptors and adhesion ligands [209]. It has been applied to quantitatively analyze the parameters of cell interactions with both 2- and 3-dimensional adhesion substrates [209],[219]. It is widely used to study intermolecular interactions [205],[222],[223]. The number of bonds between RGD and integrin receptors can be estimated in a 3D synthetic ECM [224]. Functionalization of the flagelliform peptide with organic dyes and RGD-ligands allows the estimation of force across single integrin receptors [218]. A FRET–based biosensor can follow the distribution of RhoA and Rac1 activities during cell–cell adhesion [225]. FRET can be used to explore the role of mechanical forces across proteins including actin-binding proteins and cell adhesion molecules like cadherin, PECAM-1 [216] and vinculin [50] in cellular systems. This tool may help to develop appropriate synthetic matrices useful for tissue engineering or cell-based therapies [209].

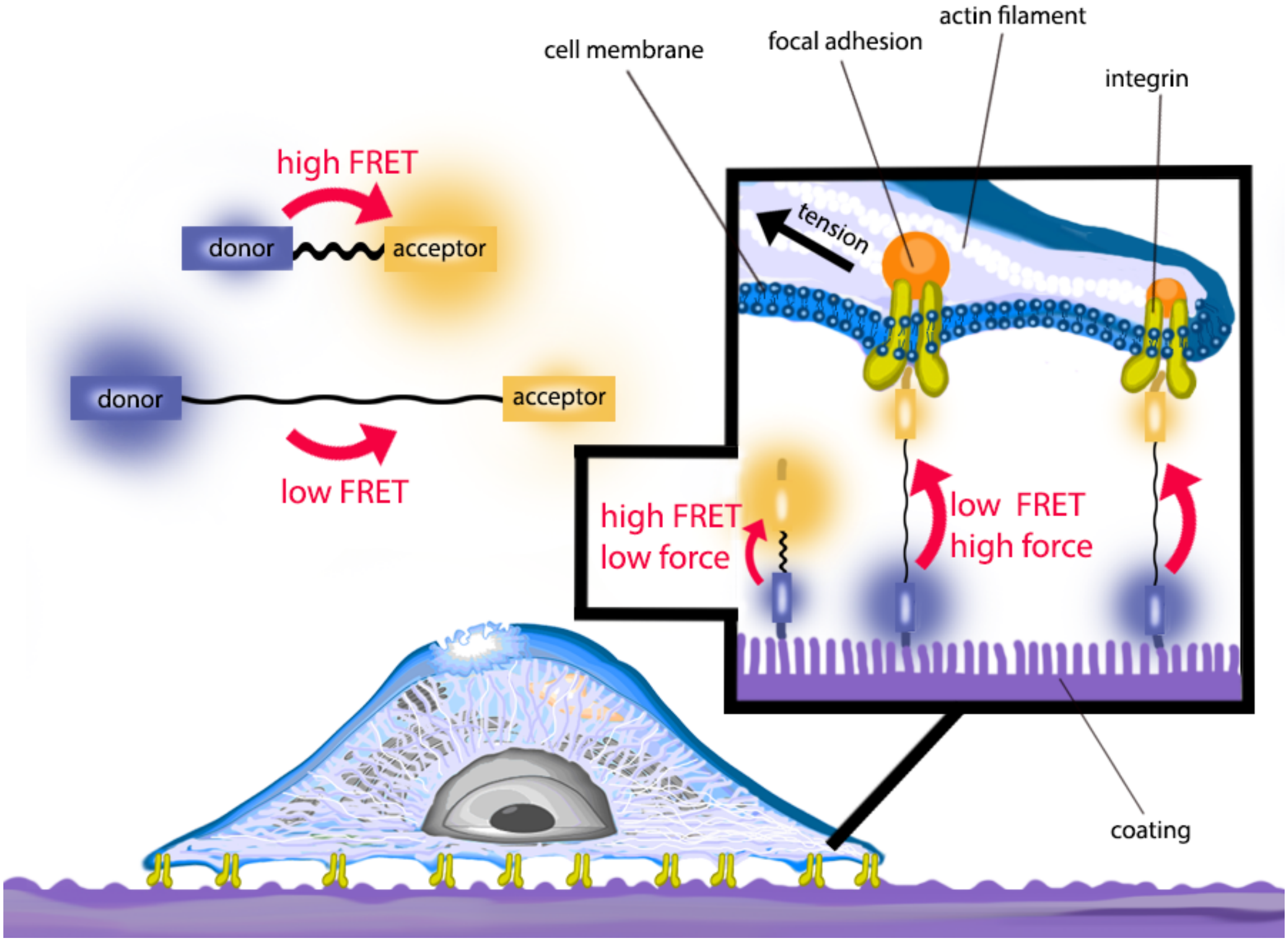

***Fig.16 Schematic illustration of the operation of FRET force sensors.*** *The method relies on the non-radiative energy transfer between two fluorophores. The emitted fluorescence of the donor molecule is quenched by the vicinity of the acceptor. Thus their distance can be calculated from the fluorescent signal. Binding a donor and an acceptor to both ends of an elastic biological polymer (such as a protein) allows to monitor the end to end distance of the polymer, which in turn characterizes the force exerted on the molecule. The technique is especially suited for the investigation of adhesion molecules such as integrins and actin binding cytoskeletal proteins.*

## Summary

Current review provides a guide to choose the appropriate technique for the measurement of cell adhesion *in vitro*. Most important parameters to be considered are the i) number of cells to be measured, ii) range of the adhesion force, iii) duration of the experiment. We identify the major advantages and disadvantages of each method. We propose that the combination of a simple cost effective method to gain data of the average behavior of a larger cell population and an advanced technique for single cell targeting can be a good strategy to obtain powerful data for biological and biophysical model development.

We also pointed out that the preparation of the adhesive surface for subsequent cellular studies has a crucial importance [22],[23]. Masking (blocking) the surface area not covered by specific biomolecules is a prerequisite for relevant results cleansed from the contribution of non-specific cell adhesion on glass, plastic or other artificial substrates [26],[30],[32],[33]. Thus we collected relevant examples and emphasized the importance of carefully controlling and pre-testing the surface chemistry.

| Method | a) Number of cells in an experiment | b) Typical range of shear stress | c) Experiment duration | d) Advantages | e) Disadvantages | f) Main applications |
|---|---|---|---|---|---|---|
| ***1.1 Centrifugal assay*** | large number of adherent cells [68] | 1-2000 pN shear stress [31],[42],[69],[70],[71] | 5-10 min [31],[42],[69],[72] | * Simplicity<br>* Widely available setup<br>* No special equipment required, only common lab instruments [42] | * Not possible to precisely determine the force acting on the cell<br>* Limited to weakly adherent cells<br>* Limited to short-term adhesion studies [42] | * Sensitive, quantitative measurement of weak binding forces [68]<br>* Quantifying cell adhesion on SAMs [69]<br>* Quantifying cell adhesion to specific RGD-modified substrates [31] |
| ***1.2 Spinning disk*** | population of cells in a single experiment [76] | 1-200 Pa shear stress [74],[77] | 5-10 min [76],[77] | * Applying a wide range of stresses in a single experiment [74]<br>* High stresses | * Not compatible with *in situ* microscopy, nor with the direct visualization of cells [61] | * Employing in more fundamental experiments [40],[75],[77] |
| ***1.3 Radial flow chamber*** | population of cells in a single experiment [42],[79],[80] | < 20 Pa shear stress [42] | 5 min [79],[81],[83],[84] | * Flexible<br>* Direct visualization of cells<br>* *In situ* observation of cell detachment [79],[83] | * Limited to weak adhesion forces [81] | * *In situ* detachment of fibroblasts [79],[81]<br>* Determination of the adhesive nature of modified substrates [79],[82],[86] |
| ***1.4 Parallel plate flow chamber*** | * 300 cells/ experimental group/ experiment [92]<br>* 100-1,000 cells/mm$^2$ [88],[91],[92] | * < 20 Pa, shear stress [42]<br>* 1-25 Pa [89],[90],[92],[93] | A few sec-210 min [89],[90],[92],[93],[94] | * Simplicity [95]<br>* Well-characterized fluid flow field [89]<br>* Direct visualization of cells [94]<br>* It can monitor cell trajectories, speed and adhesion. Motion of spherical cells in the flow can be calculated. [91] | * >200 µm channel height in most PPFC limiting the cellular adhesion force that can be measured [61] | * Determining the adhesion strength of fibroblast cells to various substrates [89]<br>* Investigation of cell retention, morphology and migration as a function of shear stress and of adhesion time [88] |
| ***1.5 Microfluidic rectangular channels*** | * ~100 cells/ experiment [102]<br>* Number of analyzed cells: 100 - 1,000 [61],[100],[101],[102] | Several hundred Pa [61],[101] | 10-15 min [61],[101] | * Direct visualization of cells [100]<br>* It can capture the cell detachment<br>* Wide range of adhesion forces can be measured<br>* High-throughput [61]<br>* Simple experimental setup | * Sustained flow affects the cytoskeleton and the morphology of cells attached to the surface. Cells become elongated in the direction of the flow being an apparent side effect of the measurement [101] | * Investigating the relationship between adhesion strength and cell geometry [101] |

***Table 1. Centrifugal and shear flow assays.***

| Method | a) Number of cells in an experiment | b) Typical force range | c) Experiment duration | d) Advantages | e) Disadvantages | f) Main applications |
|---|---|---|---|---|---|---|
| ***2.1 Step pressure technique*** | ~10 pairs of cells [109] | * Sc represents the adhesion strength; Sc =1.5 nN/μm$^2$; 744 nN separation force [46],[47] * Difficult to measure the exact contact area between the cells | * Cells are brought into contact for ~10 sec [110] * Force measurement between two cells for 120 min [47] | * Cell-cell contact area and contact time can be controlled [46],[110] * Measures cell adhesion force directly [46] | * Low throughput, manual measurement | * Measuring the adhesion force between two similar or different single cells *in vitro* [47] |
| ***2.2 Biomembrane force probe*** | 1 cell/ experiment | ❖ Vertical mode: 0.2-0.5 pN ❖ Horizon tal mode: 1-10 pN [112] | *NA* | * Adjustable loading rate of the force [46] * Sub-pN force and nanometer scale displacement resolution [46] * Gentle measurement minimally altering the cytoskeleton [111] * Can detect weak adhesive bonds | * The resolution of probe movement is not as good as in AFM [111] | * Studying receptor-ligand binding [46] |
| ***2.3 Micropipette aspiration technique*** | 1 cell/ experiment | * A force lower than 10-20 pN [46] * Hundreds of nN forces [46] | User-specified length of time (2s and 1 min) [120] | * Simplicity * Measures the force without attaching cells onto a solid surface * Can study cell-cell interactions directly * Higher forces than with optical tweezers * Versatility [46] | * Alignment of probe and cell: one of the cells (or probe) has to fit snugly inside the pipette * Cannot measure <10-20 pN forces precisely when the diameter of the pipette is ~10μm [107] * Low force sensitivity * Evaporation in the chamber is technically challenging [118] * Severe cell deformation prior to detachment | * Measuring viscoelastic properties of soft cells * Studying mechanical properties of living cells [118] |
| ***2.4 Computer controlled micropipette*** | Hundreds of cells: total number of human immune cells probed by the micropipette was 200-600 [23] | 0-2 μN [23] | ~30 min [23] | * Relatively high throughput [23], especially when compared to AFM or FluidFM * Direct visualization of cells [23] * Higher sensitivity and less side-effect than in microfluidic shear stress channel [23] | * Hydrodynamic simulations are needed to convert the experimental vacuum value to hydrodynamic lifting force [23] | * Probing single cell interactions with specific macromolecules [23] * Studying cell-cell interactions [124] |

***Table 2. Micropipette manipulation assays.***

| Method | a)<br>Number of cells in an experiment | b)<br>Typical force range | c)<br>Experiment duration | d)<br>Advantages | e)<br>Disadvantages | f)<br>Main applications |
|---|---|---|---|---|---|---|
| **3. Optical tweezers** | 1 cell/ experiment | 0.1-100 pN [127] | Manipulations of a cell takes less than 15 min [133] | * Extreme force sensitivity down to 0.1 pN [127],[130]<br><br>* High spatial resolution (0.1-2 nm) [130]<br><br>* Contactless force for manipulations [127] | * The range of applied forces is limited to max. 100 pN [130]<br><br>* Photodamage and thermal damage [127], [135] | * Interaction and binding assays [130], [132], [141],[142], [143],[144],[226]<br>* Tethered assay [127],[132], [145],[146] |

***Table 3. Optical tweezers.***

| Method | a) Number of cells in an experiment | b) Typical force range | c) Experiment duration | d) Advantages | e) Disadvantages | f) Main applications |
|---|---|---|---|---|---|---|
| **4.1 Atomic force microscopy** *4.1.1 normal force* | * a few (1-5) cell measured per condition [160] | * nN- pN forces [161]<br>* Minimum force: 50-100 pN<br>* At 100-1,000 pN force, the biological sample is reversibly deformed, above this value irreversibly deformed [163],[164],[165]<br><br>*AFM-based SCFS: 5 pN-100 nN [149] | * Chemical functionalization of the cantilever: 1h<br>* ~10 min needed for proper cell–cantilever interaction<br>* Relatively short contact times (msec-20min) [43],[170] | * Spatial and force resolution is high [148]<br>* pN force sensitivity<br>*nm positioning accuracy<br>* Can map and analyze individual receptors with nanoscale lateral resolution [148] | * Costly method [43],[48],[149]<br>* Time-consuming [161],[174]<br>* Requires a separate calibrated and functionalized cantilever for each cell [43],[48],[149]<br>* Chemical attachment of the cell to the cantilever can change the physiology of the cell [161],[174]<br>* High number of detachment force-distance curves need to be measured to gain reliable statistics [43],[170]<br>* Limited to relatively short contact times [43] | * Investigation of cell adhesion and the interactions between specific ligand-receptor pairs [149]<br>* Studying the dynamic formation of cellular adhesion<br>* Measuring cell-cell and cell-substrate adhesion [43]<br>* Extends its use from quantitatively characterizing whole-cell adhesion down to single receptor-ligand interactions [43] |
| **4.1 Atomic force microscopy** *4.1.2 Tangential/ lateral force* | 200-300 cells [2] | 10-100 nN [2],[176] | Adhesion time:10-90 min [2] | * Much lower force is sufficient to detach a particle than a normal force [177],[178] | * Interpretation of the data is not straightforward [2]<br>* The measured force depends on experimental details (shape of the tip, scanning speed, torsional cantilever spring constant) [78],[179] | * Measuring the adhesion force of protein-covered silica spheres adsorbed to polystyrene surfaces [180]<br>* Measuring the adhesion force of individual cervix cells to various substrates [2] |
| **4.2 Fluidic force microscopy** | * Up to 200 yeast and 20 mammalian cells/probe<br>* Studying ~10 cells in half an hour | 500 pN-1.6 µN [48] | * Carrying out ~10-times more experiments than conventional AFM<br>* Only a few minutes to target, immobilize, and release the cell as well as to change the cantilever position to the next cell [48] | *Directly measures the adhesion force<br>*~10 times higher throughput than conventional AFM can provide [48]<br>* Increased maximum force as compared to AFM [43]<br>* No need for chemical functionalization of the cantilever [183] | * Throughput is lower than that of the automated micropipette<br>*Microfabricated cantilevers come with a high measurement cost<br>* Cells come into direct contact with the cantilever potentially perturbing or damaging cells | * Monitoring the adhesion of single cells to specific substrates [48]<br>*Microinjection of cells [185]<br>*Microprinting [186],[187],[188] |

***Table 4. Atomic force and fluidic force microscopy.***

| Method | a) Number of cells in an experiment | b) Typical force range | c) Experiment duration | d) Advantages | e) Disadvantages | f) Main applications |
|---|---|---|---|---|---|---|
| **5.1 Cell traction force microscopy** | 1-5 in single cell experiments, several hundreds of cells in collective migration studies | * Depending on the elasticity of the substrate, which can be tuned in the physiological range [189]<br>* Not possible to resolve small forces if the overall magnitude of traction is high | * The imaging stage is only a few minutes<br>* Image analysis and deconvolution of the traction field is time consuming [193] | * Single cell measurements<br>* Remaining cells unperturbed<br>* High resolution mapping of cell-exerted forces<br>* Following the evolution of traction through time, although practically with a poor temporal resolution<br>* Using a confocal microscope, all components (x,y,z) of forces exerted on a 2D substrate can be attained (2.5D CTFM [192]<br>* Using confocal microscopy, the traction forces of a cell embedded in a 3D matrix can be determined (3D CTFM) [196] | * Extremely low throughput<br>* Image analysis and the deconvolution of the traction field is time consuming [193],[194]<br>* The traction field may be very sensitive to noises [193],[194]<br>* 2.5D and 3D CTFM requires a confocal microscope [196]<br>* 3D CTFM is in its infancy | * Mapping the 2D and 2.5D traction field of [189]<br>- individual cells [199],[200]<br>- cell aggregates [201]<br>- migrating single cells [203]<br>- a layer of collectively migrating cells [204]<br>* Mapping the traction field of cells embedded in a 3D matrix [196], [197]<br>* Correlating adhesion force with local actin density or with the size/molecular composition of adhesion complexes [191] |
| **5.2 FRET force sensors** | Minimum of 10 cells at each condition [209] | pN sensitivity [210] | ~20 min [209] | * Can detect protein-protein nanoscale interactions [219]<br>* Does not require mechanical perturbation or stimulation of cells [209]<br>* Measures the force exerted by individual molecules [220]<br>* Can be tracked in time [221] | * Artificial molecular force sensors are needed to be built into the cellular system (a specific protein)<br>* The selection of an appropriate elastic molecular element of the force sensor is critical [211],[212]<br>* Requires spectrally matched fluorophores in the force sensor [212], [220]<br>* High sensitivity camera is needed for imaging<br>* Requires data analysis algorithm to determine FRET [211] | * Exploring the role of mechanical forces built up in proteins [50],[216]<br>* Estimating the force across single integrin receptors [218]<br>* Quantifying conformational dynamics in single molecules [212]<br>* Measuring the parameters of cell interactions with 2-3D adhesion substrates [209],[219] |

***Table 5. Cell traction force microscopy and FRET force sensors.***

## Acknowledgment

### Funding sources

This work was supported by the National Research, Development and Innovation Office (grant numbers: PD 124559 for R. U. S., KH_17, KKP 129936 for R. H. ), “Lendület” Program of the Hung. Acad. Sci., ERC_HU for R. H., MedInProt grant of the Hung. Acad. Sci., Bolyai Scholarship for B. S.

### Declaration of interest

B.S. is a founder of CellSorter Company for Innovations, a startup company that developed the computer-controlled micropipette device.